\documentclass[a4paper,titlepage,12pt]{article}
\usepackage[T1]{fontenc}
\usepackage[polish,english]{babel}
\usepackage[utf8]{inputenc}
\usepackage{amsmath}
\usepackage{longtable}
\usepackage{enumerate}
\usepackage{multirow}
\usepackage{pdflscape}
\usepackage{geometry}
\usepackage{graphicx}
\usepackage{epstopdf}
\usepackage{subfig}
\usepackage{caption}
\usepackage{setspace}
\usepackage{pgfplots}
\pgfplotsset{width=13cm,height=8cm,compat=1.14}

\begin{document}

\begin{titlepage}
\begin{center}
\Large {\textbf{Soft anomalous dimension matrices in heavy quark-antiquark hadroproduction in association with a gluon jet}} 
\end{center}
\vspace{2cm} 
\begin{center} 
\Large {E. Szarek\footnote{e.szarek@th.if.uj.edu.pl}} \\[3pt]  
\normalsize{\textit{Institute of Physics, Jagiellonian University}} \\ 
\textit{Łojasiewicza 11, 30-348 Kraków, Poland} \\ [1cm] 
\end{center}

\vspace{1cm}

\begin{spacing}{2}

\section*{Abstract}

We compute the soft anomalous dimension (SAD) matrices for  production of massive quarks $Q$ and $\bar{Q}$ in association with a gluon jet, from massless quarks $q$ and antiquarks $\bar{q}$: $q\bar{q}\rightarrow Q\bar{Q}g$, and in the gluon scattering $gg\rightarrow Q\bar{Q}g$. To  analyze the behaviour of the eigenvalues of SAD matrices we perform numerical studies of their eigensystems at two special kinematical configurations.

\end{spacing}
\end{titlepage}

\clearpage

\begin{spacing}{2}

\section{Introduction}

In QCD one finds infrared divergences in perturbative corrections: soft collinear, collinear and soft non-collinear. Soft divergences appear for the energy of a gluon $E\rightarrow 0$, and the collinear ones when the angle $\theta \rightarrow 0$ between a massless parton and a gluon. After the procedure of regularization the IR divergences cancel out in observables (like cross sections), but they leave logarithmic terms depending on scales characterizing virtual and real corrections. The logarithmic remnants become very large near the absolute threshold, so they are important in processes of heavy particles production. In the absolute threshold limit the characteristic velocity of the outgoing partons $\beta$ is very small, which means that total energy $\sqrt{\hat{s}}$ of partons in center of mass system is very close to $m_{\mathrm{th}}$ where $m_{\mathrm{th}}$ is the sum of masses of products in the process. The characteristic scale of the real corrections, that come from the collinear gluon radiation is $m_{\mathrm{th}}\beta^{2}$, and the characteristic scale of the virtual corrections is proportional to $m_{\mathrm{th}}$. The real and virtual corrections combined give a leading contribution to cross section proportional to $\alpha_{s}\log^{2}\beta^{2}$. Such logarithms appear in every order of perturbative expansion contributing with $\left(\alpha_{s}\log^{2}\beta^{2}\right)^{n}$ in the leading logarithmic (LL) approximation, $\alpha_{s}^{n}\log^{2n-1}\beta^{2}$ in the next--to--leading logarithmic (NLL) approximation, and so on. When $\beta \ll 1$, $\alpha_{s}\log^{2}\beta^{2}$ may be close or grater than one, and one needs to resum those corrections to all orders. The remnant logarithms are reordered in a new perturbative expansion due to the resummation procedure. The resummation formalism is described in \cite{Collins:1985ue,kid}. The fundamental object used in the resummation procedure is the soft anomalous dimension (SAD) matrix. 
  
The soft gluon resummation technique that employs the SAD matrices have numerous applications in modern particle physics, in particular in estimates of superparticles hadroproduction. The SAD matrices carry information about colour flow between particles in the studied processes. The soft gluon resummation effects become very important in cross sections near the threshold for heavy particles production. 

The SAD matrices were calculated for various types of processes. Firstly, calculations were performed for the Drell -- Yan processes $2\rightarrow 1$ with two incoming coloured particles and one colour-neutral \cite{Sterman:1986aj,Catani:1989ne}. Then, there were considered processes $2\rightarrow 2$, like $q\bar{q}\rightarrow q\bar{q}$ and $gg\rightarrow q\bar{q}$ for massless and massive products in final state, in one-loop approximation \cite{kid}. This approach was extended to all reactions containing light quarks and gluons \cite{Oderda}. It allowed to obtain predictions for cross sections for production of heavy quarks (especially for the top quark) \cite{Bonciani:1998vc,Kidonakis:2003qe,Czakon:2009zw,Ahrens:2010zv, Cacciari:2011hy,Kidonakis:2011wy,Czakon:2013goa,Beneke:2011mq}, and compare with experimental data. The SAD matrices also play an important role in predictions for squarks and gluinos hadroproduction cross sections. Soft anomalous dimension matrices were calculated at one loop \cite{Kulesza:2008jb,Kulesza:2009kq,Beenakker:2011fu} and two loops \cite{Beenakker:2011sf,Beenakker:2014sma,Beenakker:2016gmf} for such processes. Recently a lot of effort has been made to obtain accurate predictions for reactions involving the Higgs boson. Firstly, there was obtained the hadroproduction cross section improved by the soft gluon resummation at the $\mathrm{NNLL}$ approximation \cite{Catani:2003zt} and then at the $\mathrm{N^{3}LL}$ level \cite{Bonvini:2016frm} for the $2\rightarrow 1$ process: $gg\rightarrow H^{0}$. For the supersymmetric charged Higgs boson hadroproduction the soft gluon resummation was performed at two loops for the process $bg\rightarrow tH^{-}$ \cite{Kidonakis:2010ux}. Next the soft gluon resummation was extended to a new class of processes: $2\rightarrow 3$ containing 4 coloured and 1 colour neutral particles, which gives more accurate predictions for the Higgs boson hadroproduction cross section in association with the top and antitop quarks \cite{Kulesza:2015vda,Broggio:2015lya,Broggio:2016lfj,Kulesza:2017ukk}. 

In this paper the SAD matrices are derived for $2\rightarrow 3$ processes with 5 coloured particles at one loop in the perturbation theory, $q\bar{q}\rightarrow Q\bar{Q}g$ and $gg\rightarrow Q\bar{Q}g$. The quark and antiquark in the final state are both massive. Earlier calculations of the SAD matrices in similar reactions have been performed by M. Sj\"odahl \cite{baza,colormath}, however only for massless final state protons.

\section{General Formalism}

In this paper we consider the following scattering processes:

\begin{eqnarray}
q^{\alpha}(p_{1})\bar{q}^{\beta}(p_{2})\rightarrow Q^{\gamma}(p_{3})\bar{Q}^{\delta}(p_{4})g^{a}(p_{5}),
\end{eqnarray}

\noindent and

\begin{eqnarray}
g^{a}(p_{1})g^{b}(p_{2})\rightarrow Q^{\alpha}(p_{3})\bar{Q}^{\beta}(p_{4})g^{c}(p_{5}),
\end{eqnarray}

\noindent where $\alpha$, $\beta$, $\gamma$, $\delta$, $a$, $b$ and $c$ stand for colour indices (Greek letters are used for description of a fundamental representation of $\mathrm{SU}(N_{\mathrm{c}})$ and Roman letters for an adjoint representation) and $p_{i}$, $i=1,\dots,5$, denote the momenta of particles.

Due to hard factorization theorems in QCD one can distingiush soft and hard part of sufficiently inclusive processes. The soft function $S_{IJ}$ fulfills the renormalization group equation:

\begin{equation}
\left(\mu\frac{\partial}{\partial\mu} + \beta(g)\frac{\partial}{\partial g}\right) S_{IJ} = -\left(\Gamma_{S}^{\dagger}\right)_{IB} S_{BJ} - S_{IA} \left(\Gamma_{S}\right)_{AJ},
\end{equation}

\noindent where $S_{IJ}$ is a matrix in colour space and carries information about soft wide angle gluon emissions, indices $I$, $J$ corresponds to colour tensors constructed from $\mathrm{SU}(N_{\mathrm{c}})$ representations. They depend on a studied proccess: the colour charges of participating particles and the exchange channel. For example if we consider a quark-antiquark annihilitation, $I$ and $J$ tensors correspond to a flow of a colour singlet or octet in the s-channel. At one loop, the soft anomalous dimension matrix are defined as follows \cite{kid,Oderda}:

\begin{equation}
\Gamma^{\prime}_{S}\left(g\right) = -\frac{g}{2}\frac{\partial}{\partial g}\mathrm{Res}_{\epsilon\rightarrow 0}Z_{S}\left(g,\epsilon\right),
\label{eq:gamma}
\end{equation}

\noindent where  $g$ is the coupling constant for QCD, $Z_{S}\left(g,\epsilon\right)$ -- a renormalization matrix of the soft matrix $S_{IJ}$. $Z_{S}$ receives contributions from the soft gluons. The general for of the SAD matrix can be derived from the paper
presented in \cite{Becher:2009cu,Becher:2009qa,Becher:2009kw}. However, in our explicit 
calculations we apply the method elaborated in \cite{kid,Oderda}.

 To get $Z_{S}$ one needs to sum over the contributions $Z_{S}^{(D)}$ of relevant Feynman diagrams. Each contribution $Z_{S}^{(D)}$ to $Z_{S}$ coming from a single Feynman diagram $D$, can be factorized into a colour factor and a kinematic factor:
\begin{equation}
Z_{S}^{(D)} \propto \mathrm{colour\hspace{3pt} factor} \times \mathrm{kinematic\hspace{3pt} factor}. 
\end{equation} 

\noindent The colour part of  every  diagram is represented by $\mathrm{SU}(N_{\mathrm{c}})$ tensors decomposed in an orthogonal and normalized basis. The vectors from colour basis are connected with a soft gluon line which is represented by colour tensor $if_{abc}$. The form of $Z_{S}$ depends on wheather the partons between which there is an exchange of the soft gluon, are massive or massless (see figure 1).
\noindent For massive particles $i$ and $j$ \cite{kid,Oderda}:
\begin{equation}
Z_{S}^{(D)}\left(g,\epsilon\right) = c^{ij}s_{ij}\frac{\alpha}{\pi}\frac{1}{\epsilon}\left(L_{\beta}^{(ij)} + L_{i} + L_{j} - 1\right).
\end{equation} 

\noindent For a massive particle $i$ and a massless particle $j$:

\begin{equation}
Z_{S}^{(D)}\left(g,\epsilon\right) = -c^{ij}s_{ij}\frac{\alpha}{2\pi}\frac{1}{\epsilon}\left(\ln\left[\frac{v^{2}_{ij}s}{2m_{i}^{2}}\right] - L_{i} -\ln\nu_{j} +1\right).
\end{equation}

\noindent For massless particles $i$ and $j$:

\begin{equation}
Z_{S}^{(D)}\left(g,\epsilon\right) = -c^{ij}s_{ij}\frac{\alpha}{\pi}\frac{1}{\epsilon}\left(\ln\left[\frac{\delta_{i}\delta_{j}v_{i}\cdot v_{j}}{2}\right] -\frac{1}{2}\ln\left(\nu_{i}\nu_{j}\right) +1 \right).
\end{equation}

\end{spacing}

\begin{figure}[htbp]
\centering
\includegraphics[scale=0.75]{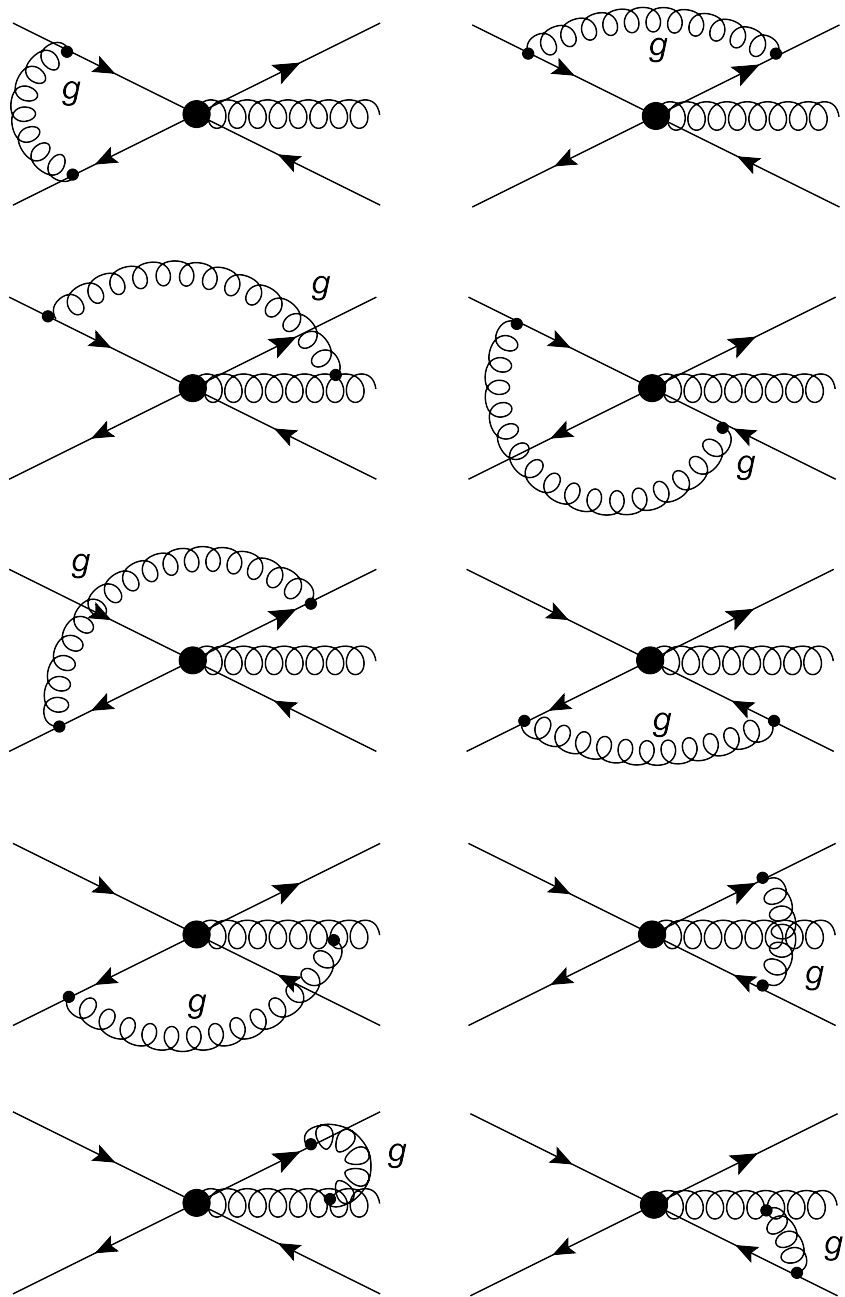}
\caption{Feynman diagrams contributing to $Z_{S}$ for process $q\bar{q}\rightarrow Q\bar{Q}g$ (for $gg\rightarrow Q\bar{Q}g$ the topologies are analogous). The soft gluon is indicated by $g$ in the diagrams.}
\end{figure} 

\begin{spacing}{2}

\noindent In the above equations factors $c^{ij}$ stand for colour factors, the number $s_{ij}$ is related to the type of particles and the direction of the momentum flow in a diagram. Namely:
\begin{equation}
s_{ij} = \Delta_{i}\Delta_{j}\delta_{i}\delta_{j}.
\end{equation}

\noindent The factors $\Delta_{i}$ depend on the type of particles between which the exchange of the gluon occurs, they have values: $+1(-1)$ for a quark (antiquark). The factors $\delta_{i} = +1(-1)$, for the same (opposite) direction of momentum flow between a parton and the soft gluon. Vectors $v_{i}$ are rescaled momenta of the particles $v^{\mu}_{i} = \frac{p^{\mu}_{i}}{Q}$, where $Q = \sqrt{\frac{\hat{s}}{2}}$, and $v_{ij} = v_{i}\cdot v_{j}$. The factors $\nu_{i} = \frac{\left(v_{i}\cdot n\right)^{2}}{|n|^{2}}$ depend on a choice of the reference vector $n^{\mu}$ of the axial gauge. In the axial gauge $A^{0}=0$ in the center of mass system of the colliding partons one has $\nu_{i} = \frac{1}{2}$. The function $L^{(ij)}_{\beta}$ depends on the relative velocity $\beta_{ij}$ of the outgoing partons:

\begin{equation}
L_{\beta}^{(ij)} = \frac{1-2m^{2}/\hat{s}_{ij}}{\beta_{ij}}\left(\ln\frac{1-\beta_{ij}}{1+\beta_{ij}} + i\pi\right),
\end{equation}

\noindent where $\beta_{ij} = \sqrt{1-4m^{2}/\hat{s}_{ij}}$ and $\hat{s}_{ij} = \left(p_{i} + p_{j}\right)^{2}$. In the processes considered the massive particles are labelled by $3$ and $4$, hence in what follows $\beta_{34}$ will be used. $L_{i}$ are dependent on the choice of gauge: $L_{i} = \frac{1}{2}\left[L_{i}(+n) + L_{i}(-n)\right]$,
\noindent where 
\begin{eqnarray}
L_{i}(\pm n) &=& \frac{1}{2}\frac{|v_{i}\cdot n|}{\sqrt{\left(v_{i}\cdot n\right)^{2} - 2m^{2}n^{2}/s}}\\ \nonumber
& & \left[\ln\left(\frac{\delta\left(\pm n\right)2m^{2}/s - |v_{i}\cdot n| - \sqrt{\left(v_{i}\cdot n\right)^{2} - 2m^{2}n^{2}/s}}{\delta\left(\pm n\right)2m^{2}/s - |v_{i}\cdot n| + \sqrt{\left(v_{i}\cdot n\right)^{2} - 2m^{2}n^{2}/s}}\right) \right.\\ \nonumber
 & & \left. + \ln\left(\frac{\delta\left(\pm n\right)n^{2} - |v_{i}\cdot n| - \sqrt{\left(v_{i}\cdot n\right)^{2} - 2m^{2}n^{2}/s}}{\delta\left(\pm n\right)n^{2} - |v_{i}\cdot n| + \sqrt{\left(v_{i}\cdot n\right)^{2} - 2m^{2}n^{2}/s}}\right)\right].
\end{eqnarray}

\noindent Contributions $L_{i}$ also appear in the self-interaction terms for the heavy quarks (antiquarks). The contribution from the self-interaction of heavy quarks (antiquarks) is

\begin{equation}
\frac{\alpha_{s}}{\pi}T_{R}\frac{N^{2}_{c} - 1}{N_{c}}\left(L_{i} + L_{j} - 2\right)\mathbf{1},
\end{equation}

where $\mathbf{1}$ is the identity matrix in the colour space and the factor $T_{R}$ comes from the normalization of generators and it equals $\frac{1}{2}$. The contribution from the self-interaction of heavy quarks (antiquarks) is added to the soft anomalous dimension matrix and the dependence of $\Gamma^{\prime}_{S}$ on $L_{i}$ is canceled out. Following refs. \cite{kid,Oderda} the Drell -- Yan contribution is subtracted from the soft anomalous dimension matrix. At one loop the Drell -- Yan SAD matrix takes the form $\frac{\alpha_{s}}{\pi}C_{F}\mathbf{1}$ ($\frac{\alpha_{s}}{\pi}C_{A}\mathbf{1}$) for the partons in the colour triplet (octet) state and  $C_{A} = T_{R}2N_{c}$ and $C_{F} = T_{R}\frac{N^{2}_{c} - 1}{N_{c}}$. The final form of the soft anomalous dimension matrix $\Gamma_{S}\left(g,\epsilon\right)$ is the following:

\begin{equation}
\Gamma_{S}\left(g,\epsilon\right) = \Gamma^{\prime}_{S}\left(g,\epsilon\right) + \frac{\alpha_{s}}{\pi}T_{R}\frac{N^{2}_{c} - 1}{N_{c}}\left(L_{3} + L_{4} - 2\right)\mathbf{1} - \frac{1}{2} \sum_{i} C^{i}_{A,F}\mathbf{1},
\label{eq:odj}
\end{equation}

\noindent where $i$ - all massless particles in the examined process.

In this paper processes with five interacting particles are considered (see figure 2). 

\end{spacing}

\begin{figure}[htbp]
\centering
\includegraphics[scale=0.75]{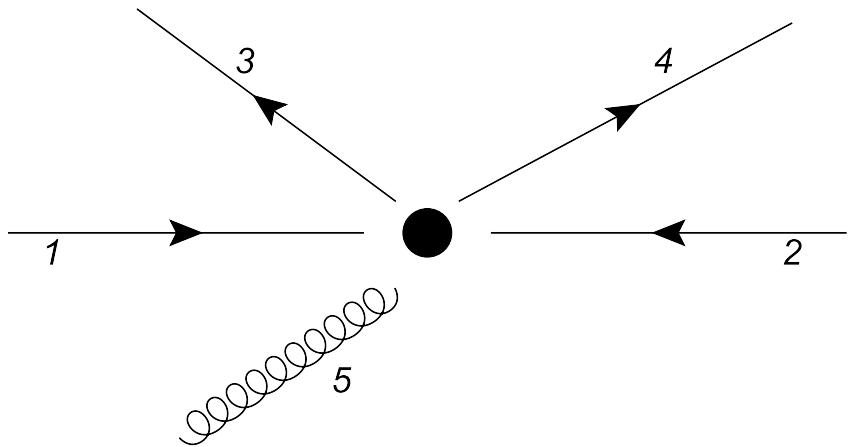}
\caption{An example of a particle collision in a process $2\rightarrow 3$.}
\end{figure} 

\begin{spacing}{2}

To fully describe the phase space of such physical system one needs five independent variables: the global azimuthal angle $\phi$, which carries  information about the rotation symmetry of the reaction and four Mandelstam--type variables:

\begin{eqnarray}
 t_{1} &=& \left(p_{3}-p_{1}\right)^{2}, \\ \nonumber
 t_{2} &=& \left(p_{4}-p_{2}\right)^{2}, \\ \nonumber
 u_{1} &=& \left(p_{3}-p_{2}\right)^{2}, \\ \nonumber
 u_{2} &=& \left(p_{4}-p_{1}\right)^{2}. 
\end{eqnarray} 

\noindent The remaining scalar products of particle momenta $p_{i}\cdot p_{5}$ $(i=1,2,3,4)$ may be expressed in the terms of above variables:

\begin{eqnarray}
p_{1}\cdot p_{5} &=& \frac{1}{2}\left(t_{1}+u_{2} + s - m^{2}_{3} - m^{2}_{4}\right), \\ \nonumber
p_{2}\cdot p_{5} &=& \frac{1}{2}\left(t_{2}+u_{1} + s - m^{2}_{3} - m^{2}_{4}\right), \\ \nonumber
p_{3}\cdot p_{5} &=& \frac{1}{2}\left(t_{2}+u_{2} + s - m^{2}_{3} - m^{2}_{4}\right), \\ \nonumber
p_{4}\cdot p_{5} &=& \frac{1}{2}\left(t_{1}+u_{1} + s - m^{2}_{3} - m^{2}_{4}\right),
\end{eqnarray}

\noindent where $m_{3,4}$ is the mass of heavy quark (antiquark).

\section{Results}

In this section we collect results for the soft anomalous dimension matrices for two processes $q\bar{q}\rightarrow Q\bar{Q}g$ and $gg\rightarrow Q\bar{Q}g$, where $q$, $\bar{q}$ -- denote the massless quark/antiquark, and $Q$, $\bar{Q}$ -- the massive quark/antiquark. Calculations of the colour factors were obtained in the $s$-channel basis, using the package ColorMath \cite{colormath} for Mathematica. The colour factors were combined with formulas (\ref{eq:gamma}), (6) -- (8), also (\ref{eq:odj}), and $\Gamma_{S}$ was obtained. It is convenient to introduce new variables $T_{1}$, $T_{2}$, $U_{1}$, $U_{2}$ and then $\Lambda$, $\Omega$, $\Gamma$, $\Sigma$, which are defined in the following way:

\begin{eqnarray}
T_{1} &=& \ln\left(\frac{2p_{1}\cdot p_{3}}{m_{3}\sqrt{s}}\right) - \frac{1 - i\pi}{2}, \\ \nonumber
T_{2} &=& \ln\left(\frac{2p_{2}\cdot p_{4}}{m_{4}\sqrt{s}}\right) - \frac{1 - i\pi}{2}, \\ \nonumber
U_{1} &=& \ln\left(\frac{2p_{2}\cdot p_{3}}{m_{4}\sqrt{s}}\right) - \frac{1 - i\pi}{2}, \\ \nonumber
U_{2} &=& \ln\left(\frac{2p_{1}\cdot p_{4}}{m_{3}\sqrt{s}}\right) - \frac{1 - i\pi}{2}.
\end{eqnarray}

\noindent In this study the case is considered of the mass of quark and antiquark that have the same value $m=m_{3}=m_{4}$ and we introduce variables,

\begin{eqnarray}
\Lambda &=& T_{1}+T_{2}+U_{1}+U_{2},\\ \nonumber
\Omega &=& T_{1}+T_{2}-U_{1}-U_{2}, \\ \nonumber
\Gamma &=& T_{1}-T_{2}+U_{1}-U_{2}, \\ \nonumber
\Sigma &=& T_{1}-T_{2}-U_{1}+U_{2} .
\end{eqnarray}

\noindent In what follows, the SAD matrices will be presented in terms of the independent variables $\Lambda$, $\Omega$, $\Gamma$, $\Sigma$ and the variables $v_{i5}=2p_{i}\cdot p_{5}/s$ that can be expressed by $\Lambda$, $\Omega$, $\Gamma$, $\Sigma$ with equalities (11), (12) and (13). The variables $v_{i5}$ are kept in order to simplify the form of the matrices.

\subsection{$q\bar{q}\rightarrow Q\bar{Q}g$}
 
The following orthogonal and normalized colour basis was used in the calculations \cite{baza}:
\begin{eqnarray}
T^{1}_{\alpha\beta\gamma\zeta a} &=& \frac{1}{\sqrt{N_{c}\left(N^{2}_{c}-1\right)T_{R}}}\delta_{\alpha\beta}t^{a}_{\gamma\zeta}, \\ \nonumber
T^{2}_{\alpha\beta\gamma\zeta a} &=& \frac{1}{\sqrt{N_{c}\left(N^{2}_{c}-1\right)T_{R}}}\delta_{\gamma\zeta}t^{a}_{\beta\alpha}, \\ \nonumber
T^{3}_{\alpha\beta\gamma\zeta a} &=& \frac{1}{\sqrt{2N_{c}\left(N^{2}_{c}-1\right)\left(T_{R}\right)^{3}}}t^{b}_{\beta\alpha}t^{c}_{\gamma\zeta}if_{bca}, 
\end{eqnarray}

\begin{equation*}
T^{4}_{\alpha\beta\gamma\zeta a} = \frac{\sqrt{N_{c}}}{\sqrt{2\left(N^{2}_{c}-4\right)\left(N^{2}_{c}-1\right)\left(T_{R}\right)^{3}}}t^{b}_{\beta\alpha}t^{c}_{\gamma\zeta}d_{bca}. 
\end{equation*}

\noindent Notice that indices of the adjoint representation are $a$, $b$, $c$ and indices of the fundamental representation are $\alpha$, $\beta$, $\gamma$ and $\zeta$. The following results are valid for $N_{c}\geq 3$ \cite{baza}.
 
The soft anomalous dimension matrix $\Gamma_{q\bar{q}\rightarrow Q\bar{Q}g}$ can be split into two parts:

\begin{equation}
\Gamma_{q\bar{q}\rightarrow Q\bar{Q}g} = \Gamma^{(1)}_{q\bar{q}\rightarrow Q\bar{Q}g}\left(\Lambda, \Omega, \Gamma, \Sigma\right) + \Gamma^{(2)}_{q\bar{q}\rightarrow Q\bar{Q}g}\left( v_{i5}\right),
\end{equation}

\noindent where $\Gamma^{(1)}$ receives contributions from the soft gluon exchanges between particles $1,2,3,4$, and $\Gamma^{(2)}$ from exchanges between particles $i$ and $5$, with $i=1,2,3,4$. Hence:

\end{spacing}

\begin{equation}
\begin{split}
&\Gamma^{(1)}_{q\bar{q}\rightarrow Q\bar{Q}g} = \frac{\alpha_{s}}{\pi}T_{R} \times\\
& \left(\begin{array}{cccc}
\frac{1+\left(1-N_{c}^{2}\right) L_{\beta}}{N_{c}} & \frac{\Omega}{N_{c}} & \frac{\sqrt{N_{c}^{2} - 4}}{\sqrt{2}N_{c}}\Omega & \frac{\Gamma}{\sqrt{2}}\\
\frac{\Omega}{N_{c}} & \frac{2+2L_{\beta} + N_{c}^{2}}{2N_{c}} & \frac{\sqrt{N_{c}^{2} - 4}}{\sqrt{2}N_{c}}\Omega & \frac{\Sigma}{\sqrt{2}} \\
\frac{\sqrt{N_{c}^{2} - 4}}{\sqrt{2}N_{c}}\Omega & \frac{\sqrt{N_{c}^{2} - 4}}{\sqrt{2}N_{c}}\Omega & \frac{4+4L_{\beta} + \left(N_{c}^{2}-12\right)\Omega + N_{c}^{2}\left(2 + \Lambda\right)}{4N_{c}} & \frac{1}{4}\sqrt{N_{c}^{2}-4}\left(\Gamma+\Sigma\right) \\
\frac{\Gamma}{\sqrt{2}}  &  \frac{\Sigma}{\sqrt{2}} & \frac{1}{4}\sqrt{N_{c}^{2}-4}\left(\Gamma+\Sigma\right) &  \frac{4+4L_{\beta} + \left(N_{c}^{2}-4\right)\Omega + N_{c}^{2}\left(2 + \Lambda\right)}{4N_{c}}
\end{array}\right),
\end{split}
\end{equation}

\noindent and

\begin{small}
\begin{equation}
\begin{split}
& \Gamma^{(2)}_{q\bar{q}\rightarrow Q\bar{Q}g} = \frac{\alpha_{s}}{\pi}T_{R} \times\\
&\left(\begin{array}{cccc}
N_{c}\ln\left(v_{15}v_{25}\right) & 0 & 0 & \frac{1}{\sqrt{2}}\ln\left(\frac{v_{45}}{v_{35}}\right)\\
0 & \frac{1}{2}N_{c}\ln\left(v_{35}v_{45}\right) & 0 & \ln\left(\frac{v_{25}}{v_{15}}\right)\sqrt{2} \\
0 & 0 & \frac{1}{4}N_{c}\ln\left(v^{2}_{15}v^{2}_{25}v_{35}v_{45}\right) & \frac{1}{4}\ln\left(\frac{v^{2}_{25}v_{45}}{v^{2}_{15}v_{35}}\right)\sqrt{N_{c}^{2}-4} \\
\frac{1}{\sqrt{2}}\ln\left(\frac{v_{45}}{v_{35}}\right)  &  \ln\left(\frac{v_{25}}{v_{15}}\right)\sqrt{2} & \frac{1}{4}\ln\left(\frac{v^{2}_{25}v_{45}}{v^{2}_{15}v_{35}}\right)\sqrt{N_{c}^{2}-4} & \frac{1}{4}N_{c}\ln\left(v^{2}_{15}v^{2}_{25}v_{35}v_{45}\right)
\end{array}\right) \\
& \vspace{6pt} \\
& +\quad \frac{\alpha_{s}}{\pi}T_{R}\times\mathbf{diag}\left(\frac{i\pi}{N_{c}},\: i\pi\frac{1-N^{2}_{c}}{N_{c}},\: i\pi\frac{2-N^{2}_{c}}{2N_{c}},\: i\pi\frac{2-N^{2}_{c}}{2N_{c}}\right).
\end{split}
\end{equation}
\end{small}

\begin{spacing}{2}

\noindent The SAD matrices for $q\bar{q}\rightarrow Q\bar{Q}g$ were calculated in parallel in \cite{Schaefer}\footnote{Several differences were found between results of \cite{Schaefer} and the results of this paper and their origin was clarified in correspondence with R. Sch\"afer.}. Note that, the obtained SAD matrix is complex symmetric. This 
property has been proved to hold in general in an orthonormal basis 
\cite{Seymour:2008xr}. The same feature will be found also for the $gg$-channel. For clarity we denoted $L^{(34)}_{\beta}$ as $L_{\beta}$ in all the matrices. The general form of the soft anomalous dimension matrix is rather complicated, hence in order to provide more insight into its properties we consider special kinematical configurations for which the matrix simplifies. First, we consider the case when the momenta of the outgoing quark and antiquark are equal, $p^{\mu}_{3}=p^{\mu}_{4}$. Then variables $\Lambda$, $\Omega$, $\Gamma$, $\Sigma$ reduce to $\Lambda\rightarrow \Lambda^{\prime} = 2T_{1}+2U_{1}$, $\Sigma\rightarrow \Sigma^{\prime} = 2T_{1}-2U_{1}$, $\Gamma\rightarrow 0$, $\Omega\rightarrow 0$. In this special case it is convenient to introduce a variable $\beta = \sqrt{1 - \frac{4m^{2}}{\hat{s}}}$. In this limit the form of the soft anomalous matrix becomes significantly simpler: 

\end{spacing}

\begin{equation}
\begin{split}
& \Gamma^{(1)}_{q\bar{q}\rightarrow Q\bar{Q}g}\left(p^{\mu}_{3}=p^{\mu}_{4}\right) = \frac{\alpha_{s}}{\pi}T_{R}\times \\
&\left(\begin{array}{cccc}
\frac{1+\left(1-N_{c}^{2}\right) L_{\beta}}{N_{c}} & 0 & 0 & 0\\
0 & \frac{2+2L_{\beta} + N_{c}^{2}}{2N_{c}} & 0 & \frac{\Sigma^{\prime}}{\sqrt{2}} \\
0 & 0 & \frac{4+4L_{\beta} + N_{c}^{2}\left(2 + \Lambda^{\prime}\right)}{4N_{c}} & \frac{\Sigma^{\prime}}{4}\sqrt{N_{c}^{2}-4} \\
0  &  \frac{\Sigma^{\prime}}{\sqrt{2}} & \frac{\Sigma^{\prime}}{4}\sqrt{N_{c}^{2}-4} & \frac{4+4L_{\beta} + N_{c}^{2}\left(2 + \Lambda^{\prime}\right)}{4N_{c}}
\end{array}\right)
\end{split}
\end{equation}

\begin{small}
\begin{equation}
\begin{split}
& \Gamma^{(2)}_{q\bar{q}\rightarrow Q\bar{Q}g}\left(p^{\mu}_{3}=p^{\mu}_{4}\right) = \frac{\alpha_{s}}{\pi}T_{R}\times \\
&\left(\begin{array}{cccc}
N_{c}\ln\left(v_{15}v_{25}\right) & 0 & 0 & 0\\
0 & N_{c}\ln v_{35} & 0 & \ln\left(\frac{v_{25}}{v_{15}}\right)\sqrt{2} \\
0 & 0 & \frac{1}{2} N_{c}\ln\left(v_{15}v_{25}v_{35}\right) & \frac{1}{2}\ln\left(\frac{v_{25}}{v_{15}}\right)\sqrt{N_{c}^{2}-4} \\
0 & \ln\left(\frac{v_{25}}{v_{15}}\right)\sqrt{2} & \frac{1}{2}\ln\left(\frac{v_{25}}{v_{15}}\right)\sqrt{N_{c}^{2}-4} & \frac{1}{2}N_{c}\ln\left( v_{15}v_{25}v_{35}\right)
\end{array}\right) \\
& \vspace{6pt} \\
& +\quad \frac{\alpha_{s}}{\pi}T_{R}\times\mathbf{diag}\left(\frac{i\pi}{N_{c}},\: i\pi\frac{1-N^{2}_{c}}{N_{c}},\: i\pi\frac{2-N^{2}_{c}}{2N_{c}},\: i\pi\frac{2-N^{2}_{c}}{2N_{c}}\right).
\end{split}
\end{equation}
\end{small}

\begin{spacing}{2}

\noindent It can be seen that the soft anomalous dimension matrix can be divided in two blocks: $\mathbf{1\times 1}$ and $\mathbf{3\times 3}$.

Next the limit $\Sigma^{\prime}=0$ is performed that corresponds to $p_{1}\cdot p_{3} = p_{2}\cdot p_{3}$. The obtained matrix has a diagonal form:

\end{spacing}

\begin{equation}
\begin{split}
&\Gamma_{q\bar{q}\rightarrow Q\bar{Q}g}\left(\Sigma^{\prime}=0\right) = \frac{\alpha_{s}}{\pi}T_{R}\times \\
&\vspace{6pt} \\
&\left\{ \frac{1}{N_{c}}\times \mathbf{diag}\left(1+\left(1 - N_{c}^{2}\right)L_{\beta},\: 1+L_{\beta} + \frac{N^{2}_{c}}{2},\: 1+L_{\beta} + \frac{N^{2}_{c}}{4}\left(2 + \Lambda^{\prime}\right),\: 1+L_{\beta} + \frac{N^{2}_{c}}{4}\left(2 + \Lambda^{\prime}\right)\right) \right.\\
&\left. \vspace{6pt}  \right.\\
&\left. +\quad \mathbf{diag}\left(2N_{c}\ln v_{15},\: N_{c}\ln v_{35},\: \frac{1}{2}N_{c}\ln\left(v^{2}_{15}v_{35}\right),\: \frac{1}{2}N_{c}\ln\left(v^{2}_{15}v_{35}\right)\right) \right.\\
&\left. \vspace{6pt} \right.\\
&\left. +\quad \mathbf{diag}\left(\frac{i\pi}{N_{c}},\: i\pi\frac{1-N^{2}_{c}}{N_{c}},\: i\pi\frac{2-N^{2}_{c}}{2N_{c}},\: i\pi\frac{2-N^{2}_{c}}{2N_{c}}\right)\right\}.
\end{split}
\end{equation}

\begin{spacing}{2}

\subsubsection{Analysis of the eigenvalues for $q\bar{q}\rightarrow Q\bar{Q}g$}

In this section we consider the behaviour of the SAD eigenvalues for $p_{3}=p_{4}$ and two different scattering angles $\theta$ ($90^{\circ}$ and $30^{\circ}$) where $\theta$ is an angle between the incoming and outgoing partons in the CMS frame. $\theta = 90^{\circ}$ represents the most symmetrical case, and the choice of $\theta = 30^{\circ}$ represents a less symmetrical configuration. The limit $\Sigma^{\prime}\rightarrow 0$ corresponds to the case of $\theta = 90^{\circ}$. This analysis must be done carrefully  because $L^{(34)}_{\beta}$ in the limit $\beta_{34}\rightarrow 0$ gives singular terms $\propto \frac{i}{\beta_{34}}$. We need to execute three steps. While performing the limit $\beta_{34}\rightarrow 0$ one subtracts the singular terms from the SAD matrix and then the limit $\beta \rightarrow 0$ may be studied. Finally we present results after a subtraction of the asymptotic small $\beta$ behaviour that is treated analytically. Numerical calculations were performed for the $N_{c} = 3$ case. The eigenvalues of $\tilde{\Gamma}_{S}$ does not contain the prefactor $\frac{\alpha_{s}}{\pi}$. The relation between the full SAD matrix $\Gamma_{S}$ and $\tilde{\Gamma}_{S}$ is $\Gamma_{S} = \frac{\alpha_{s}}{\pi}\tilde{\Gamma}_{S}$. The singular matrix subtracted from $\tilde{\Gamma}_{S}$ has a form $\frac{i\pi}{\beta_{34}}\times\mathbf{diag}\left(-\frac{2}{3},\frac{1}{12},\frac{1}{12},\frac{1}{12}\right)$.

For a general $\beta$ there is a degeneracy of eigenvalues for $\theta = 90^{\circ}$, there are three different eigenvalues. For $\theta = 30^{\circ}$ there is no degeneracy. All eigenvalues are complex with non-trival real and imaginary part. For $\theta=90^{\circ}$ there are two different values of the imaginary part of eigenvalues instead of four, which is the case of $30^{\circ}$.

For $\beta\rightarrow 0$ one finds a singular term proportional to $\log \beta$ which gives a contribution to the real part of the eigenvalues. Each eigenvalue of $\Gamma_{S}$ has the same leading behaviour in $\beta \rightarrow 0$ for both scattering angles. One finds one asymptotic form of the eigenvalues of small $\beta$: 

\begin{equation}
\lambda^{\mathrm{sing}} = 6\log\beta.
\end{equation}

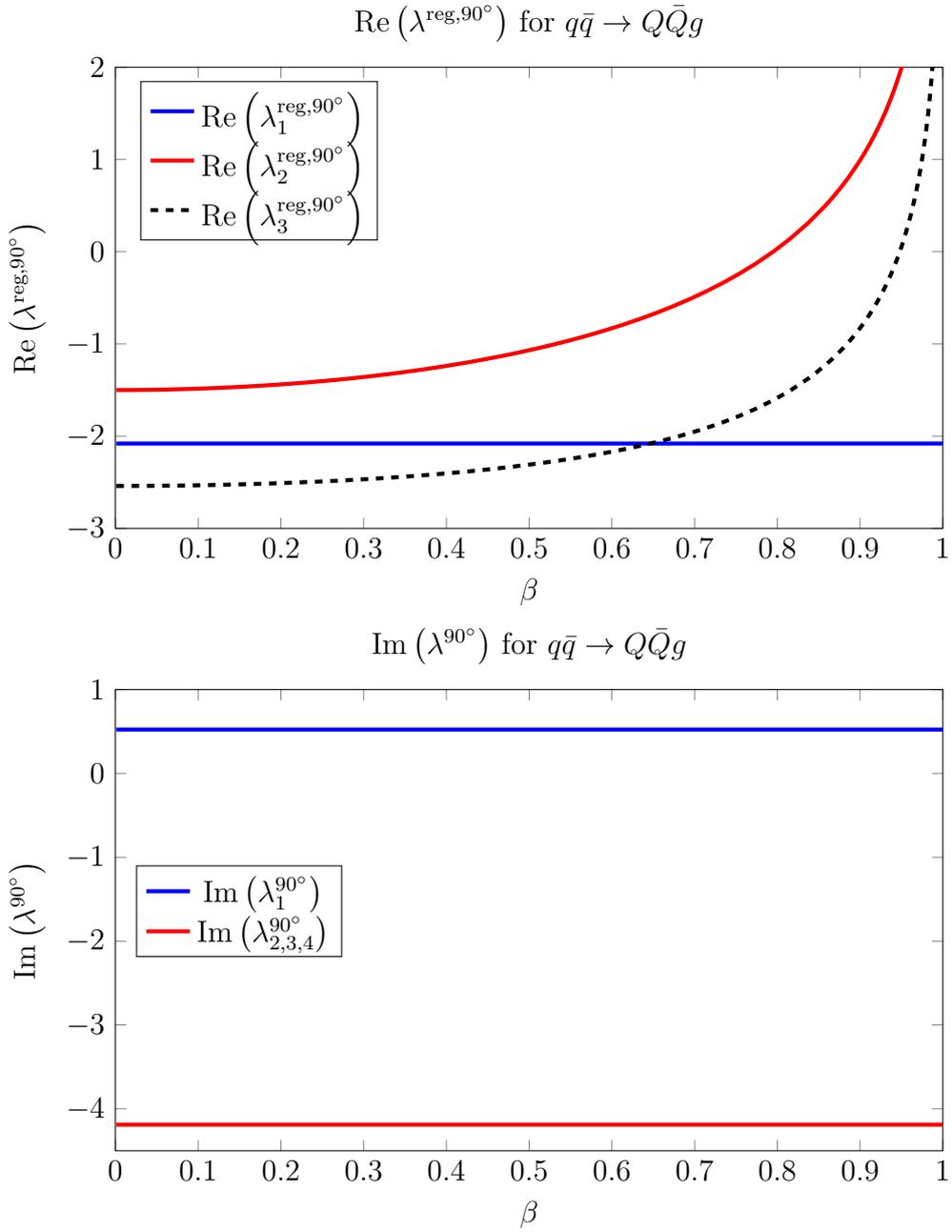
\begin{figure}
\centering
\begin{tikzpicture}
\begin{axis}[title=$\mathrm{Re}\left(\lambda^{\mathrm{reg},90^{\circ}}\right)$ for $q\bar{q}\rightarrow Q\bar{Q}g$,
xlabel={$\beta$},
ylabel={$\mathrm{Re}\left(\lambda^{\mathrm{reg},90^{\circ}}\right)$},
ymin=-3, ymax=2,
xmin=0, xmax=1,
legend pos=north west,
legend entries={$\mathrm{Re}\left(\lambda^{\mathrm{reg},90^{\circ}}_{1}\right)$,
$\mathrm{Re}\left(\lambda^{\mathrm{reg},90^{\circ}}_{2}\right)$,
$\mathrm{Re}\left(\lambda^{\mathrm{reg},90^{\circ}}_{3}\right)$},]
\addplot[blue, ultra thick] table {kwarkidrell90re1.dat};
\addplot[red, ultra thick] table {kwarkidrell90re2.dat};
\addplot[black, dashed, ultra thick] table {kwarkidrell90re3.dat};
\end{axis}
\end{tikzpicture}
\\
\begin{tikzpicture}
\pgfplotsset{every axis legend/.append style={at={(0.025,0.42)},anchor=south west}}
\begin{axis}[title=$\mathrm{Im}\left(\lambda^{90^{\circ}}\right)$ for $q\bar{q}\rightarrow Q\bar{Q}g$,
xlabel={$\beta$},
ylabel={$\mathrm{Im}\left(\lambda^{90^{\circ}}\right)$},
ymin=-4.5, ymax=1,
xmin=0, xmax=1,
legend entries={$\mathrm{Im}\left(\lambda^{90^{\circ}}_{1}\right)$,
$\mathrm{Im}\left(\lambda^{90^{\circ}}_{2,3,4}\right)$},]
\addplot[blue, ultra thick] table {kwarki90ima1.dat};
\addplot[red, ultra thick] table {kwarki90im2.dat};
\end{axis}
\end{tikzpicture}
\caption{The real (top) and imaginary (bottom) parts of the eigenvalues of $\tilde{\Gamma}_{S}$ \qquad for $q\bar{q}\rightarrow Q\bar{Q}g$ at $\theta = 90^{\circ}$.}
\label{fig:kwarki90}
\end{figure}

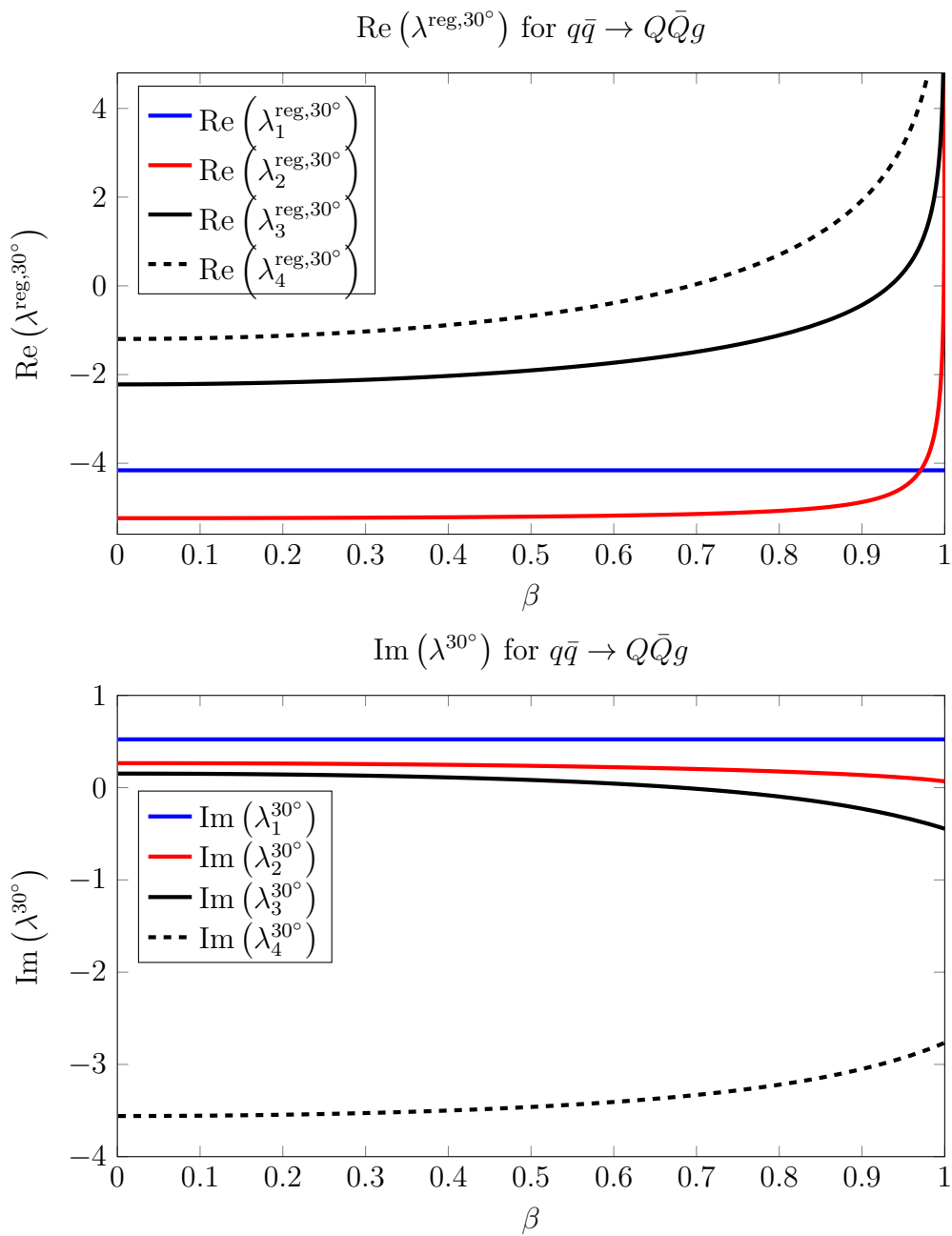
\begin{figure}
\centering
\begin{tikzpicture}
\pgfplotsset{every axis legend/.append style={at={(0.025,0.52)},anchor=south west}}
\begin{axis}[title=$\mathrm{Re}\left(\lambda^{\mathrm{reg},30^{\circ}}\right)$ for $q\bar{q}\rightarrow Q\bar{Q}g$,
xlabel={$\beta$},
ylabel={$\mathrm{Re}\left(\lambda^{\mathrm{reg},30^{\circ}}\right)$},
ymin=-5.6, ymax=4.8,
xmin=0, xmax=1,
legend entries={$\mathrm{Re}\left(\lambda^{\mathrm{reg},30^{\circ}}_{1}\right)$,
$\mathrm{Re}\left(\lambda^{\mathrm{reg},30^{\circ}}_{2}\right)$,
$\mathrm{Re}\left(\lambda^{\mathrm{reg},30^{\circ}}_{3}\right)$,
$\mathrm{Re}\left(\lambda^{\mathrm{reg},30^{\circ}}_{4}\right)$},]
\addplot[blue, ultra thick] table {kwarkidrell30re1.dat};
\addplot[red, ultra thick] table {kwarkidrell30re2.dat};
\addplot[black, ultra thick] table {kwarkidrell30re3.dat};
\addplot[black, dashed, ultra thick] table {kwarkidrell30re4.dat};
\end{axis}
\end{tikzpicture}
\\
\begin{tikzpicture}
\pgfplotsset{every axis legend/.append style={at={(0.025,0.42)},anchor=south west}}
\begin{axis}[title=$\mathrm{Im}\left(\lambda^{30^{\circ}}\right)$ for $q\bar{q}\rightarrow Q\bar{Q}g$,
xlabel={$\beta$},
ylabel={$\mathrm{Im}\left(\lambda^{30^{\circ}}\right)$},
ymin=-4, ymax=1,
xmin=0, xmax=1,
legend entries={$\mathrm{Im}\left(\lambda^{30^{\circ}}_{1}\right)$,
$\mathrm{Im}\left(\lambda^{30^{\circ}}_{2}\right)$,
$\mathrm{Im}\left(\lambda^{30^{\circ}}_{3}\right)$,
$\mathrm{Im}\left(\lambda^{30^{\circ}}_{4}\right)$},]
\addplot[blue, ultra thick] table {kwarki30ima1.dat};
\addplot[red, ultra thick] table {kwarki30im2.dat};
\addplot[black, ultra thick] table {kwarki30im3.dat};
\addplot[black, dashed, ultra thick] table {kwarki30im4.dat};
\end{axis}
\end{tikzpicture}
\caption{The real (top) and imaginary (bottom) parts of the eigenvalues of $\tilde{\Gamma}_{S}$ \qquad for $q\bar{q}\rightarrow Q\bar{Q}g$ at $\theta = 30^{\circ}$.}
\label{fig:kwarki30}
\end{figure}

In Figures \ref{fig:kwarki90} and \ref{fig:kwarki30} we show regularized eigenvalues. They are defined as $\lambda^{\mathrm{reg}}_{i} = \lambda_{i} - \lambda^{\mathrm{sing}}$. One observes a quite similar behaviour of the regularized eigensystem for $\theta = 90^{\circ}$ and $\theta = 30^{\circ}$. In Figure \ref{fig:kwarki90} one can see that all $\mathrm{Re}(\lambda^{\mathrm{reg},90^{\circ}})$ are either constant ($\lambda_{1}$) or slightly increasing ($\lambda_{2}$,$\lambda_{34}$) up to $\beta\approx 0.4$. $\mathrm{Im}(\lambda^{90^{\circ}})$ are constant in $\beta$. In the case of  $\theta = 30^{\circ}$, the real and imaginary part of $\lambda_{1}$ shows a constant behaviour. The real parts of $\lambda_{2}$, $\lambda_{3}$ and $\lambda_{4}$ exhibit a similar behaviour for small $\beta$ -- they are slowly varying for moderate $\beta$, then for $\beta >0.6$ they are rapidly increasing.  The imaginary parts of these remaining eigenvalues are nearly constant for $\beta <0.7$, and then start to slowly decrease ($\lambda_{2}$ and $\lambda_{3}$) or slowly increase ($\lambda_{4}$).
\end{spacing}

\pagebreak
\begin{spacing}{2}

\subsection{$gg\rightarrow Q\bar{Q}g$}

The following orthogonal and normalized colour basis was used in the calculations \cite{baza}: \end{spacing}

\begin{eqnarray}
T^{1}_{ab\alpha\beta c} &=& \frac{1}{\left(N^{2}_{c}-1\right)\sqrt{T_{R}}}t^{c}_{\alpha\beta}\delta_{ab}, \\ \nonumber
T^{2}_{ab\alpha\beta c} &=& \frac{1}{N_{c}\sqrt{2\left(N^{2}_{c}-1\right)T_{R}}}if_{abc} \delta_{\alpha\beta}, \\ \nonumber
T^{3}_{ab\alpha\beta c} &=& \frac{1}{\sqrt{2\left(N^{2}_{c}-4\right)\left(N^{2}_{c}-1\right)T_{R}}}d_{abc} \delta_{\alpha\beta}, \\ \nonumber
T^{4}_{ab\alpha\beta c} &=& \frac{1}{2N_{c}\sqrt{\left(N^{2}_{c}-1\right)\left(T_{R}\right)^{3}}}if_{abn}if_{mcn}t^{m}_{\alpha\beta}, \\ \nonumber
T^{5}_{ab\alpha\beta c} &=& \frac{1}{\sqrt{4\left(N^{2}_{c}-4\right)\left(N^{2}_{c}-1\right)\left(T_{R}\right)^{3}}} d_{abn}if_{mcn}t^{m}_{\alpha\beta}, \\  \nonumber
T^{6}_{ab\alpha\beta c} &=& \frac{1}{\sqrt{4\left(N^{2}_{c}-4\right)\left(N^{2}_{c}-1\right)\left(T_{R}\right)^{3}}} if_{abn}d_{mcn}t^{m}_{\alpha\beta}, \\ \nonumber
T^{7}_{ab\alpha\beta c} &=& \frac{1}{\sqrt{4\left(N^{2}_{c}-4\right)^{2}\left(N^{2}_{c}-1\right)\left(T_{R}\right)^{3}}} d_{abn}d_{mcn}t^{m}_{\alpha\beta}, \\  \nonumber
T^{8}_{ab\alpha\beta c} &=& \frac{1}{\sqrt{2\left(N^{2}_{c}-4\right)\left(N^{2}_{c}-1\right)\left(T_{R}\right)^{3}}} P^{10+\bar{10}}_{abmc}t^{m}_{\alpha\beta}, \\ \nonumber
T^{9}_{ab\alpha\beta c} &=& \frac{1}{\sqrt{2\left(N^{2}_{c}-4\right)\left(N^{2}_{c}-1\right)\left(T_{R}\right)^{3}}} P^{10-\bar{10}}_{abmc}t^{m}_{\alpha\beta}, \\ \nonumber
T^{10}_{ab\alpha\beta c} &=& \frac{-1}{\sqrt{N^{2}_{c}\left(N_{c}+3\right)\left(N_{c}-1\right)\left(T_{R}\right)^{3}}} P^{27}_{abmc}t^{m}_{\alpha\beta}, \\ \nonumber
T^{11}_{ab\alpha\beta c} &=& \frac{1}{\sqrt{N^{2}_{c}\left(N_{c}-3\right)\left(N_{c}+1\right)\left(T_{R}\right)^{3}}} P^{0}_{abmc}t^{m}_{\alpha\beta},
\end{eqnarray}

\noindent where 

\begin{eqnarray}
P^{10+\bar{10}}_{abcd} &=& \frac{1}{2}\left(\delta_{ac}\delta_{bd}-\delta_{ad}\delta_{cb}\right) - \frac{1}{N_{c}}f_{abg}f_{cdg}, \\ \nonumber 
P^{10-\bar{10}}_{abcd} &=& \frac{1}{2}d_{acg}if_{bgd} - \frac{1}{2}d_{bgd}if_{acg}, \\ \nonumber
P^{27}_{abcd} &=& \frac{N_{c}}{4\left(N_{c}+2\right)}d_{abg}d_{cdg} + \frac{1}{2}f_{adg}f_{cbg} - \frac{1}{4}f_{abg}f_{cdg} + \frac{1}{4}\delta_{ad}\delta_{bc} + \frac{1}{4}\delta_{ac}\delta_{bd} \\ \nonumber
& & + \frac{1}{2\left(N_{c}+1\right)}\delta_{ab}\delta_{cd}, \\ \nonumber
P^{0}_{abcd} &=& - \frac{N_{c}}{4\left(N_{c}+2\right)}d_{abg}d_{cdg} - \frac{1}{2}f_{adg}f_{cbg} + \frac{1}{4}f_{abg}f_{cdg} + \frac{1}{4}\delta_{ad}\delta_{bc} + \frac{1}{4}\delta_{ac}\delta_{bd} \\ \nonumber
& & - \frac{1}{2\left(N_{c}+1\right)}\delta_{ab}\delta_{cd}. \\ \nonumber
\end{eqnarray}

\begin{spacing}{2}
As for the $q\bar{q}\rightarrow Q\bar{Q}g$ case the soft anomalous dimension matrix $\Gamma_{gg\rightarrow Q\bar{Q}g}$ is split in two parts:

\begin{equation}
\Gamma_{gg\rightarrow Q\bar{Q}g} = \Gamma^{(1)}_{gg\rightarrow Q\bar{Q}g}\left(\Lambda, \Omega, \Gamma, \Sigma\right) + \Gamma^{(2)}_{gg\rightarrow Q\bar{Q}g}\left(v_{i5}\right),
\end{equation}

\noindent where
\end{spacing}

\newpage
\newgeometry{tmargin=2.5cm, bmargin=2.5cm, lmargin=1cm, rmargin=1cm} 

\begin{landscape}
\begin{footnotesize}
\begin{equation}
\begin{split}
& \Gamma^{(1)}_{gg\rightarrow Q\bar{Q}g} = \frac{\alpha_{s}}{\pi}T_{R}\times \\
& \left(\begin{array}{ccccccc}
\frac{1+\left(1-N_{c}^{2}\right)L_{\beta}}{N_{c}} & -\Omega & 0 & -\frac{\Omega}{\sqrt{2}} & \frac{\Gamma}{\sqrt{2}} & 0 & \dots\\ 
 -\Omega & \frac{2+2L_{\beta} +N_{c}^{2}}{N_{c}} & 0 & 0 & 0 & \frac{\Gamma}{2} & \dots\\ 
 0 & 0 & \frac{1+\left(1-N_{c}^{2}\right)L_{\beta}}{N_{c}}  & 0 & 0 & 0 & \dots \\
 -\frac{\Omega}{\sqrt{2}} & 0 & 0 & \frac{4+4L_{\beta} +N_{c}^{2}\left(2+\Lambda\right)}{4N_{c}} & -\frac{\Sigma N_{c}}{4} & 0 & \dots\\ 
 \frac{\Gamma}{\sqrt{2}} & 0 & 0 & -\frac{\Sigma N_{c}}{4} &  \frac{4+4L_{\beta} +N_{c}^{2}\left(2+\Lambda\right)}{4N_{c}} & -\frac{\Omega}{\sqrt{2}} & \dots\\ 
 0 & \frac{\Gamma}{2} & 0 & 0 & \frac{-\Omega}{\sqrt{2}} & \frac{2+2L_{\beta} + N_{c}^{2}\left(2 + \Lambda\right)}{2N_{c}} & \dots\\ 
 0 & \frac{\Omega}{2\sqrt{2}} \sqrt{\frac{\left(N_{c} - 3\right)\left(N_{c} - 1\right)\left(N_{c} + 2\right)}{\left(N_{c} - 2\right)}} & -\Omega \sqrt{\frac{N_{c} + 1}{2\left(N_{c} - 1\right)}} & \frac{\Omega}{2} \sqrt{\frac{\left(N_{c} - 3\right)\left(N_{c} + 2\right)}{\left(N_{c} - 2\right)\left(N_{c} - 1\right)}} & 0 & \frac{\Sigma}{2\sqrt{2}} \sqrt{\frac{\left(N_{c} - 3\right)\left(N_{c} - 1\right)\left(N_{c} + 2\right)}{\left(N_{c} - 2\right)}} & \dots\\ 
0 & -\frac{\Omega\sqrt{2}}{\sqrt{N^{2}_{c}-4}} & \frac{-\Omega}{\sqrt{2}} & -\frac{\Omega\left(N_{c}^{2}-12\right)}{4\sqrt{N_{c}^{2} - 4}} & \frac{\Gamma}{4}\sqrt{N_{c}^{2}-4} & \frac{N_{c}\Sigma}{\sqrt{2}\sqrt{N_{c}^{2} - 4}} & \dots\\
0 & \frac{\Omega}{2\sqrt{2}}\sqrt{\frac{\left(N_{c}+3\right)\left(N_{c}+1\right)\left(N_{c}-2\right)}{\left(N_{c}+2\right)}} & -\Omega\sqrt{\frac{N_{c}+3}{2\left(N_{c}+1\right)}} & \frac{-\Omega}{2}\sqrt{\frac{\left(N_{c}-2\right)\left(N_{c}+3\right)}{\left(N_{c}+1\right)\left(N_{c}+2\right)}} & 0 & \frac{-\Sigma}{2\sqrt{2}}\sqrt{\frac{\left(N_{c}-2\right)\left(N_{c}+3\right)}{\left(N_{c}+1\right)\left(N_{c}+2\right)}}  & \dots\\ 
0 & 0 & \frac{\Gamma}{\sqrt{2}} & \frac{\Gamma}{4}\sqrt{N_{c}^{2} - 4} & \frac{-\Gamma}{4}\sqrt{N_{c}^{2} - 4} & 0 & \dots\\ 
0 & 0 & \frac{-\Omega\sqrt{2}}{\sqrt{N^{2}_{c}-1}} & \Omega\sqrt{\frac{N_{c}^{2} - 4}{N_{c}^{2} - 1}} & 0 & 0 & \dots 
\end{array}\right. \\
& \vspace{12pt} \\
& \left. \begin{array}{cccccc}
\dots & 0 & 0 & 0 & 0 & 0\\
\dots & \frac{\Omega}{2\sqrt{2}} \sqrt{\frac{\left(N_{c} - 3\right)\left(N_{c} - 1\right)\left(N_{c} + 2\right)}{\left(N_{c} - 2\right)}} & -\frac{\Omega\sqrt{2}}{\sqrt{N^{2}_{c}-4}} & \frac{\Omega}{2\sqrt{2}}\sqrt{\frac{\left(N_{c}+3\right)\left(N_{c}+1\right)\left(N_{c}-2\right)}{\left(N_{c}+2\right)}}  & 0 & 0\\
\dots & -\Omega \sqrt{\frac{N_{c} + 1}{2\left(N_{c} - 1\right)}} & \frac{-\Omega}{\sqrt{2}} & -\Omega\sqrt{\frac{N_{c}+3}{2\left(N_{c}+1\right)}} & \frac{\Gamma}{\sqrt{2}} & \frac{-\Omega\sqrt{2}}{\sqrt{N^{2}_{c}-1}}\\
\dots & \frac{\Omega}{2}\sqrt{\frac{\left(N_{c}-3\right)\left(N_{c}+2\right)}{\left(N_{c}-2\right)\left(N_{c}-1\right)}}  & -\frac{\Omega\left(N^{2}_{c}-12\right)}{4\sqrt{N^{2}_{c}-4}} & -\frac{\Omega}{2}\sqrt{\frac{\left(N_{c}+3\right)\left(N_{c}-2\right)}{\left(N_{c}+2\right)\left(N_{c}+1\right)}} & \frac{-\Gamma}{4}\sqrt{N_{c}^{2} - 4} & \Omega\sqrt{\frac{N_{c}^{2} - 4}{N_{c}^{2} - 1}}\\
\dots & 0 & \frac{\Gamma}{4}\sqrt{N^{2}_{c}-4} & 0 & \frac{-\Omega}{4}\sqrt{N^{2}_{c}-4} & 0 \\
\dots & -\frac{\Sigma N_{c}}{4} & \frac{\Sigma}{2}\sqrt{\frac{N_{c}+3}{N_{c}+1}} & -\frac{\Sigma}{2}\sqrt{\frac{N_{c}-3}{N_{c}-1}} & 0 & 0 \\
\dots & \frac{4+4L_{\beta} + N_{c}^{2}\left(3 + \Lambda\right)}{4N_{c}} & 0 & 0 & \frac{\Sigma N_{c}}{\sqrt{2\left(N^{2}_{c}-4\right)}} & \frac{\Omega\sqrt{2}}{\sqrt{N^{2}_{c}-4}}\\
\dots & 0 & \frac{2+2L_{\beta} + N_{c}\left[2N_{c}+1 +\Lambda\left(N_{c}+1\right)\right]}{2N_{c}} & 0 & -\frac{\Sigma}{2}\sqrt{\frac{\left(N_{c}-2\right)\left(N_{c}+1\right)\left(N_{c}+3\right)}{2\left(N_{c}-2\right)}} & \frac{\Omega}{2}\sqrt{\frac{\left(N_{c}-2\right)\left(N_{c}+1\right)\left(N_{c}+3\right)}{2\left(N_{c}-2\right)}}\\
\dots & 0 & 0 & \frac{2+2L_{\beta} + N_{c}\left(N_{c}+1\right)\left(2 +\Lambda\right)}{2N_{c}} & \frac{\Sigma}{2}\sqrt{\frac{N_{c}+3}{N_{c}+1}} & 0\\
\dots & \frac{-\Sigma}{2}\sqrt{\frac{N_{c} - 3}{N_{c} - 1}} & \frac{-N_{c}\Sigma}{4} & \frac{\Sigma}{2}\sqrt{\frac{N_{c} + 3}{N_{c} + 1}} & \frac{4 + 4L_{\beta} + N_{c}^{2}\left(2 + \Lambda\right)}{4N_{c}} & \frac{-N_{c}\Sigma}{\sqrt{N_{c}^{2} -1}}\\
\dots & 0 & 0 & 0 &  \frac{-N_{c}\Sigma}{\sqrt{N_{c}^{2} -1}} & \frac{1 + L_{\beta} }{N_{c}}
\end{array}\right),
\end{split}
\end{equation}
\end{footnotesize}
\end{landscape}

\newpage

\begin{landscape}
\begin{footnotesize}
\begin{equation}
\begin{split}
& \Gamma^{(2)}_{gg\rightarrow Q\bar{Q}g} = \frac{\alpha_{s}}{\pi}T_{R}\times \\
& \left(\begin{array}{cccccccc}
N_{c}\ln\left(v_{15}v_{25}\right) & 0 & 0 & 0 & \frac{1}{\sqrt{2}}\ln\left(\frac{v_{45}}{v_{35}}\right) & 0 & 0 & \dots \\ 
 0 & N_{c}\ln\left(v_{15}v_{25}\right) & 0 & 0 & 0 & \frac{1}{2}\ln\left(\frac{v_{45}}{v_{35}}\right) & 0 & \dots \\ 
 0 & 0 & N_{c}\ln\left(v_{15}v_{25}\right) & 0 & 0 & 0 & 0 & \dots \\ 
0 & 0 & 0 & \frac{1}{4}N_{c}\ln\left(v^{2}_{15}v^{2}_{25}v_{35}v_{45}\right) & \frac{1}{2}N_{c}\ln\left(\frac{v_{15}}{v_{25}}\right) & 0 & 0 & \dots \\ 
 \frac{1}{\sqrt{2}}\ln\left(\frac{v_{45}}{v_{35}}\right) & 0 & 0 & \frac{1}{2}N_{c}\ln\left(\frac{v_{15}}{v_{25}}\right) &  \frac{1}{4}N_{c}\ln\left(v^{2}_{15}v^{2}_{25}v_{35}v_{45}\right) & 0 & 0 & \dots \\  
 0 & \frac{1}{2}\ln\left(\frac{v_{45}}{v_{35}}\right) & 0 & 0 & 0 & N_{c}\ln\left(v_{15}v_{25}\right) &   & \dots \\ 
 0 & 0 & 0 & 0 & 0 &  & 2\left(N_{c} - 1\right)\ln\left(v_{15}v_{25}\right) + \ln\left(v_{35}v_{45}\right) & \dots \\ 
0 & 0 & 0 & 0 & \frac{1}{4}\ln\left(\frac{v_{45}}{v_{35}}\right)\sqrt{N^{2}_{c}-4} & \frac{N_{c}\ln\left(\frac{v_{25}}{v_{15}}\right)\sqrt{2}}{\sqrt{N^{2}_{c}-4}} & 0 & \dots  \\
0 & 0 & 0 & 0 & 0 & \ln\left(\frac{v_{15}}{v_{25}}\right)\sqrt{\frac{\left(N_{c} + 3\right)\left(N_{c} + 1\right)\left(N_{c} - 2\right)}{N_{c}+2}} & 0 & \dots  \\
0 & 0 & \frac{1}{\sqrt{2}}\ln\left(\frac{v_{45}}{v_{35}}\right) & \frac{\ln\left(\frac{v_{25}}{v_{15}}\right)}{4}\sqrt{N^{2}_{c}-4} & 0 & 0 & \ln\left(\frac{v_{25}}{v_{15}}\right)\sqrt{\frac{N_{c}-3}{{N}_{c}-1}} & \dots \\ 
0 & 0 & 0 & 0 & 0 & 0 & 0 & \dots   
\end{array}\right. \\
& \vspace{12pt} \\
& \left. \begin{array}{ccccc}
\dots  & 0 & 0 & 0 & 0\\
\dots  & 0 & 0 & 0 & 0\\
\dots  & 0 & 0 & \ln\left(\frac{v_{45}}{v_{35}}\right)\frac{1}{\sqrt{2}} & 0\\
\dots  & 0 & 0 & \frac{1}{4}\ln\left(\frac{v_{45}}{v_{35}}\right)\sqrt{N^{2}_{c}-4} & 0\\
\dots  & \frac{1}{4}\ln\left(\frac{v_{45}}{v_{35}}\right)\sqrt{N^{2}_{c}-4} & 0 & 0 & 0 \\
\dots  & \frac{\ln\left(\frac{v_{25}}{v_{15}}\right)N_{c}\sqrt{2}}{\sqrt{N^{2}_{c}-4}} & \ln\left(\frac{v_{15}}{v_{25}}\right) \sqrt{\frac{\left(N_{c}+3\right)\left(N_{c}+1\right)\left(N_{c}-2\right)}{N_{c}+2}}\frac{1}{\sqrt{2}} & 0 & 0\\
\dots  & 0 & 0 & \ln\left(\frac{v_{15}}{v_{25}}\right)\sqrt{\frac{N_{c}-3}{{N}_{c}-1}} & 0\\
\dots  & \frac{N_{c}}{4}\ln\left(v^{2}_{15}v^{2}_{25}v_{35}v_{45}\right) & 0 & \frac{N_{c}}{2}\ln\left(\frac{v_{25}}{v_{15}}\right) & 0\\
\dots  & 0 &\left(N_{c}+1\right)\ln\left(v_{15}v_{25}\right) - \frac{1}{2}\ln\left(v_{35}v_{45}\right) & \ln\left(\frac{v_{25}}{v_{15}}\right)\sqrt{\frac{N_{c}+3}{{N}_{c}+1}} & 0\\
\dots  & \frac{N_{c}}{2}\ln\left(\frac{v_{25}}{v_{15}}\right) & \ln\left(\frac{v_{25}}{v_{15}}\right)\sqrt{\frac{N_{c}+3}{{N}_{c}+1}} & \frac{N_{c}}{4}\ln\left(v^{2}_{15}v^{2}_{25}v_{35}v_{45}\right) & \frac{2N_{c}}{\sqrt{N^{2}_{c}-1}}\ln\left(\frac{v_{15}}{v_{25}}\right)\\
\dots & 0 & 0 & \frac{2N_{c}}{\sqrt{N^{2}_{c}-1}}\ln\left(\frac{v_{15}}{v_{25}}\right) &  N_{c}\ln\left(v_{35}v_{45}\right)
\end{array}\right) \\
& \vspace{12pt}  \\
& +\quad \mathbf{diag}\left(-i\pi N_{c},\: -i\pi N_{c},\: -2i\pi N_{c},\: -\frac{3}{2}i\pi N_{c},\: -\frac{3}{2}i\pi N_{c},\: -\frac{3}{2}i\pi N_{c},\: -\frac{3}{2}i\pi N_{c},\: -i\pi\left(N_{c}-1\right),\: -i\pi\left(N_{c}+1\right),\: -i\pi N_{c},\: -i\pi N_{c}\right).
\end{split}
\end{equation}
\end{footnotesize}
\end{landscape}

\newpage
\newgeometry{tmargin=2.5cm}

In the next step a special case is considered $p^{\mu}_{3}=p^{\mu}_{4}$. The obtained matrix has a block -- diagonal form:

\begin{equation}
\Gamma_{gg\rightarrow Q\bar{Q}g}\left(p^{\mu}_{3}=p^{\mu}_{4}\right) = \frac{\alpha_{s}}{\pi}T_{R}\times \left(\begin{array}{ccc}
\Gamma_{\mathbf{3\times 3}} & & \\
 & \Gamma_{\mathbf{2\times 2}} & \\
 & & \Gamma_{\mathbf{6\times 6}} 
\end{array}\right) 
\end{equation}

\noindent where

\begin{eqnarray}
\Gamma_{\mathbf{3\times 3}} & = & \frac{1}{N_{c}}\times \mathrm{\textbf{diag}}\left\{1+\left(1-N_{c}^{2}\right)L_{\beta}+N_{c}^{2}\ln\left(v_{15}v_{25}\right) - i\pi N_{c}^{2}, \right. \\ \nonumber
& & \left. 1 + L_{\beta} + N_{c}^{2}\left(1 - i\pi + \frac{\Lambda^{\prime}}{2} + \ln\left(v_{15}v_{25}\right)\right), \right. \\ \nonumber
& & \left. 1 + \left(1 - N_{c}^{2}\right)L_{\beta} + N_{c}^{2}\ln\left(v_{15}v_{25}\right) - i\pi N_{c}^{2}, \right\}, \\ \nonumber
& & \\ \nonumber
\Gamma_{\mathbf{2\times 2}} & = & \left(\begin{array}{cc}
\frac{4 + 4L_{\beta} + N_{c}^{2}\left(2 - 6i\pi + \Lambda^{\prime} + 2\ln\left(v_{15}v_{25}v_{35}\right) \right)}{4N_{c}}  & -\frac{N_{c}}{4}\left(\Sigma^{\prime} - 2\ln\left(\frac{v_{25}}{v_{15}}\right)\right) \\
-\frac{N_{c}}{4}\left(\Sigma^{\prime} - 2\ln\left(\frac{v_{25}}{v_{15}}\right)\right)  & \frac{4 + 4L_{\beta} + N_{c}^{2}\left(2 - 6i\pi + \Lambda^{\prime} + 2\ln\left(v_{15}v_{25}v_{35}\right) \right)}{4N_{c}} 
\end{array}\right), 
\end{eqnarray}

\noindent and

\begin{landscape}
\begin{footnotesize}
\begin{equation}
\begin{split}
& \Gamma_{\mathbf{6\times 6}} =  \\
& \left(\begin{array}{cccc}
 \frac{1 + L_{\beta} + N_{c}^{2}\left(1 - i\pi + \frac{\Lambda^{\prime}}{2} + \ln\left(v_{15}v_{25}\right)\right)}{N_{c}} & \frac{1}{2\sqrt{2}} \sqrt{\frac{\left(N_{c} - 3\right)\left(N_{c} - 1\right)\left(N_{c} + 2\right)}{N_{c} - 2}}\left(\Sigma^{\prime} - 2\ln\left(\frac{v_{25}}{v_{15}}\right)\right) &  \frac{N_{c}}{\sqrt{2}\sqrt{N_{c}^{2} -4}}\left(\Sigma^{\prime} - 2\ln\left(\frac{v_{25}}{v_{15}}\right)\right) & \dots \\
\frac{1}{2\sqrt{2}} \sqrt{\frac{\left(N_{c} - 3\right)\left(N_{c} - 1\right)\left(N_{c} + 2\right)}{N_{c} - 2}}\left(\Sigma^{\prime} - 2\ln\left(\frac{v_{25}}{v_{15}}\right)\right) & \frac{1 + L_{\beta} + N_{c}\left(N_{c} - 1\right)\left(1 + \Lambda^{\prime} +2\ln\left(v_{15}v_{25}\right)\right)- 2i\pi N_{c}\left(N_{c} + 1\right) + 2N_{c}\ln v_{35}}{N_{c}} & 0 & \dots \\
 \frac{N_{c}}{\sqrt{2}\sqrt{N_{c}^{2} -4}}\left(\Sigma^{\prime} - 2\ln\left(\frac{v_{25}}{v_{15}}\right)\right) & 0 & \frac{4 + 4L_{\beta} + N_{c}^{2}\left(2 - 6i\pi + \Lambda^{\prime} + 2\ln\left(v_{15}v_{25}v_{35}\right) \right)}{4N_{c}} & \dots \\
 \frac{-1}{2\sqrt{2}} \sqrt{\frac{\left(N_{c} + 3\right)\left(N_{c} + 1\right)\left(N_{c} - 2\right)}{N_{c} + 2}}\left(\Sigma^{\prime} - 2\ln\left(\frac{v_{25}}{v_{15}}\right)\right) & 0 & 0 & \dots\\
0 &  \frac{-1}{2}\sqrt{\frac{N_{c} - 3}{N_{c} - 1}}\left(\Sigma^{\prime} - 2\ln\left(\frac{v_{25}}{v_{15}}\right)\right) & \frac{-N_{c}}{4}\left(\Sigma^{\prime} - 2\ln\left(\frac{v_{25}}{v_{15}}\right)\right) & \dots \\
0 & 0 & 0 & \dots
\end{array}\right. \\
& \vspace{12pt} \\
& \left. \begin{array}{cccc}
\dots &  \frac{-1}{2\sqrt{2}} \sqrt{\frac{\left(N_{c} + 3\right)\left(N_{c} + 1\right)\left(N_{c} - 2\right)}{N_{c} + 2}}\left(\Sigma^{\prime} - 2\ln\left(\frac{v_{25}}{v_{15}}\right)\right) & 0 & 0 \\
\dots & 0 & \frac{-1}{2}\sqrt{\frac{N_{c} - 3}{N_{c} - 1}}\left(\Sigma^{\prime} - 2\ln\left(\frac{v_{25}}{v_{15}}\right)\right) & 0 \\
\dots & 0 & \frac{-N_{c}}{4}\left(\Sigma^{\prime} - 2\ln\left(\frac{v_{25}}{v_{15}}\right)\right) & 0 \\
\dots & \frac{2 + 2L_{\beta} + N_{c}\left(N_{c} + 1\right)\left(2 + \Lambda^{\prime} +2\ln\left(v_{15}v_{25}\right)\right)- 2i\pi N_{c}\left(N_{c} - 1\right) - 2N_{c}\ln v_{35}}{2N_{c}}  & \frac{1}{2}\sqrt{\frac{N_{c} + 3}{N_{c} + 1}}\left(\Sigma^{\prime} - 2\ln\left(\frac{v_{25}}{v_{15}}\right)\right) & 0 \\
\dots & \frac{1}{2}\sqrt{\frac{N_{c} + 3}{N_{c} + 1}}\left(\Sigma^{\prime} - 2\ln\left(\frac{v_{25}}{v_{15}}\right)\right) & \frac{4 + 4L_{\beta} + N_{c}^{2}\left(2 - 6i\pi + \Lambda^{\prime} + 2\ln\left(v_{15}v_{25}v_{35}\right) \right)}{4N_{c}} & \frac{-N_{c}}{\sqrt{N_{c}^{2} -1}}\left(\Sigma^{\prime} - 2\ln\left(\frac{v_{25}}{v_{15}}\right)\right) \\
\dots & 0 &  \frac{-N_{c}}{\sqrt{N_{c}^{2} -1}}\left(\Sigma^{\prime} - 2\ln\left(\frac{v_{25}}{v_{15}}\right)\right) &  \frac{1 + L_{\beta} + N_{c}^{2}\left(\ln v_{35} - 2i\pi\right)}{N_{c}}
\end{array}\right). 
\end{split}
\end{equation}
\end{footnotesize}

\vspace{12pt}
\noindent For the case $N_{c} = 3$, the last block becomes even simpler: $\Gamma_{\mathbf{6\times 6}} = \Gamma_{\mathbf{1\times 1}}\otimes \Gamma_{\mathbf{5\times 5}}$.

\noindent After performing the limit $\Sigma^{\prime}=0$ matrices take the following form:
\end{landscape}

\begin{small}
\begin{eqnarray}
\begin{split}
& \Gamma_{gg\rightarrow Q\bar{Q}g}\left(\Sigma^{\prime}=0\right) =  \frac{\alpha_{s}}{\pi}T_{R}\times \frac{1}{N_{c}}\times \\ 
&\mathrm{\textbf{diag}}\left\{1+\left(1-N_{c}^{2}\right)L_{\beta}+N_{c}^{2}\ln\left(v_{15}v_{25}\right) - i\pi N_{c}^{2}, 1 + L_{\beta} + N_{c}^{2}\left(1 - i\pi + \frac{\Lambda^{\prime}}{2} + \ln\left(v_{15}v_{25}\right)\right), \right. \\ \nonumber
 & \left. 1 + \left(1 - N_{c}^{2}\right)L_{\beta} + N_{c}^{2}\ln\left(v_{15}v_{25}\right) - i\pi N_{c}^{2}, 1 + L_{\beta} + \frac{N_{c}^{2}}{4}\left(2 - 6i\pi + \Lambda^{\prime} + 2\ln\left(v_{15}v_{25}v_{35}\right) \right), \right. \\ \nonumber
 & \left. 1 + L_{\beta} + \frac{N_{c}^{2}}{4}\left(2 - 6i\pi + \Lambda^{\prime} + 2\ln\left(v_{15}v_{25}v_{35}\right) \right), 1 + L_{\beta} + N_{c}^{2}\left(1 - i\pi + \frac{\Lambda^{\prime}}{2} + \ln\left(v_{15}v_{25}\right)\right), \right. \\ \nonumber
 & \left. 1 + L_{\beta} + N_{c}\left(N_{c} - 1\right)\left(1 + \Lambda^{\prime} +2\ln\left(v_{15}v_{25}\right)\right)- 2i\pi N_{c}\left(N_{c} + 1\right) + 2N_{c}\ln v_{35}, \right. \\ \nonumber
 & \left. 1 + L_{\beta} + \frac{N_{c}^{2}}{4}\left(2 - 6i\pi + \Lambda^{\prime} + 2\ln\left(v_{15}v_{25}v_{35}\right) \right), \right. \\ \nonumber
 & \left. 1 + L_{\beta} + \frac{N_{c}}{2}\left(N_{c} + 1\right)\left(2 + \Lambda^{\prime} +2\ln\left(v_{15}v_{25}\right)\right)- i\pi N_{c}\left(N_{c} - 1\right) - N_{c}\ln v_{35}, \right. \\ \nonumber
 & \left. 1 + L_{\beta} + \frac{N_{c}^{2}}{4}\left(2 - 6i\pi + \Lambda^{\prime} + 2\ln\left(v_{15}v_{25}v_{35}\right) \right), 1 + L_{\beta} + N_{c}^{2}\left(\ln v_{35} - 2i\pi\right)\right\}.
\end{split}
\end{eqnarray}
\end{small}

\begin{spacing}{2}

\subsubsection{Analysis of the eigenvalues for $gg\rightarrow Q\bar{Q}g$}

\begin{figure}
\centering
\begin{tikzpicture}
\begin{axis}[title=$\mathrm{Re}\left(\lambda^{\mathrm{reg},90^{\circ}}\right)$ for $gg\rightarrow Q\bar{Q}g$,
xlabel={$\beta$},
ylabel={$\mathrm{Re}\left(\lambda^{\mathrm{reg},90^{\circ}}\right)$},
ymin=-3, ymax=2,
xmin=0, xmax=1,
legend pos=north west,
legend entries={$\mathrm{Re}\left(\lambda^{\mathrm{reg},90^{\circ}}_{1,2}\right)$,
$\mathrm{Re}\left(\lambda^{\mathrm{reg},90^{\circ}}_{3}\right)$,
$\mathrm{Re}\left(\lambda^{\mathrm{reg},90^{\circ}}_{4,5,6,7}\right)$},]
\addplot[blue, ultra thick] table {gluonydrell90re1.dat};
\addplot[red, ultra thick] table {gluonydrell90re3.dat};
\addplot[black, dashed, ultra thick] table {gluonydrell90re4.dat};
\end{axis}
\end{tikzpicture}
\\
\begin{tikzpicture}
\begin{axis}[title=$\mathrm{Re}\left(\lambda^{\mathrm{reg},90^{\circ}}\right)$ for $gg\rightarrow Q\bar{Q}g$,
xlabel={$\beta$},
ylabel={$\mathrm{Re}\left(\lambda^{\mathrm{reg},90^{\circ}}\right)$},
ymin=-4.5, ymax=0,
xmin=0, xmax=1,
legend pos=north west,
legend entries={$\mathrm{Re}\left(\lambda^{\mathrm{reg},90^{\circ}}_{8}\right)$,
$\mathrm{Re}\left(\lambda^{\mathrm{reg},90^{\circ}}_{9}\right)$,
$\mathrm{Re}\left(\lambda^{\mathrm{reg},90^{\circ}}_{10}\right)$},]
\addplot[blue, ultra thick] table {gluonydrell90re8.dat};
\addplot[red, ultra thick] table {gluonydrell90re9.dat};
\addplot[black, dashed, ultra thick] table {gluonydrell90re10.dat};
\end{axis}
\end{tikzpicture}
\caption{The real parts of the regularized eigenvalues of $\tilde{\Gamma}_{S}$ for $gg\rightarrow Q\bar{Q}g$ \qquad \qquad at $\theta = 90^{\circ}$.}
\label{fig:gluonyre90}
\end{figure}
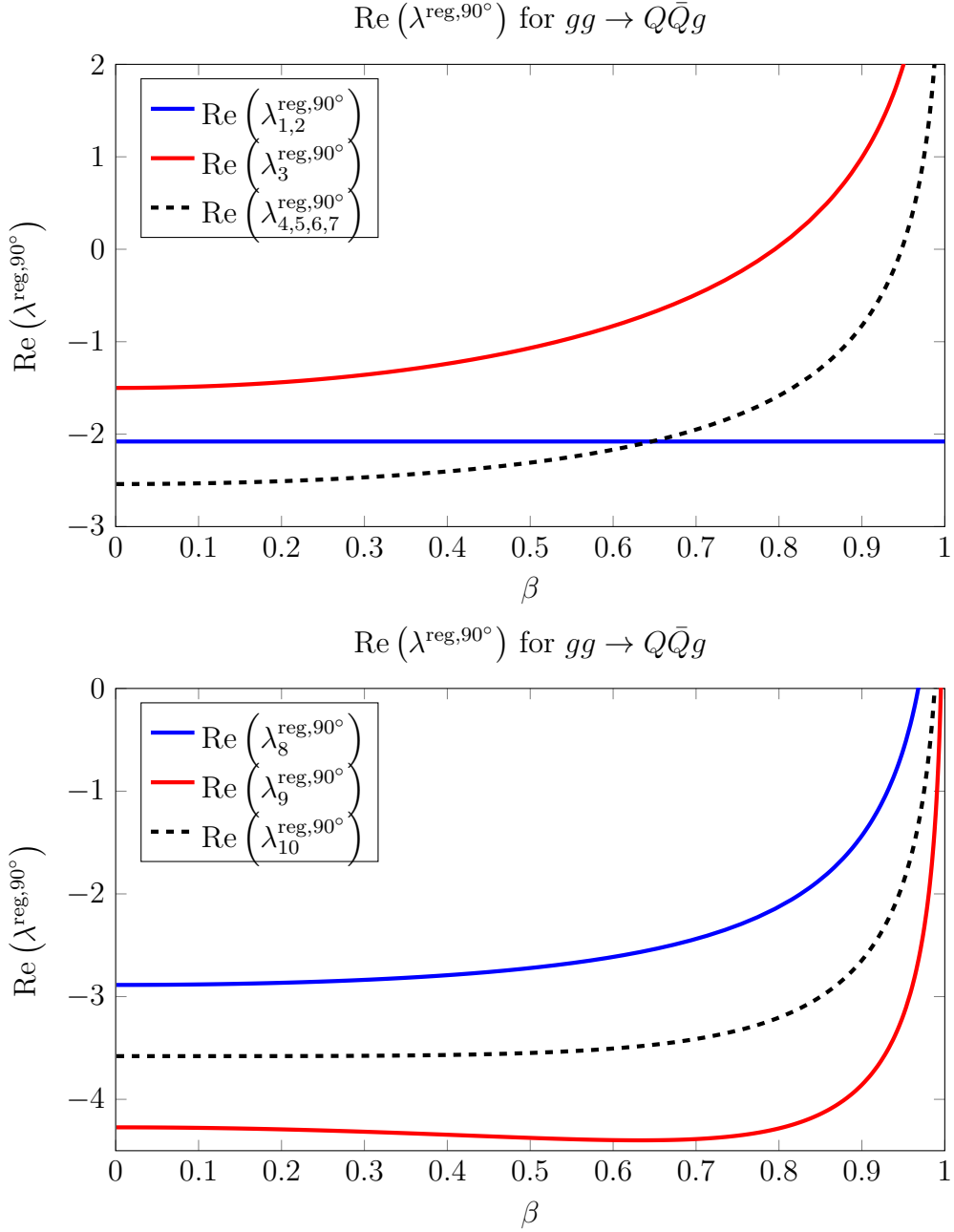

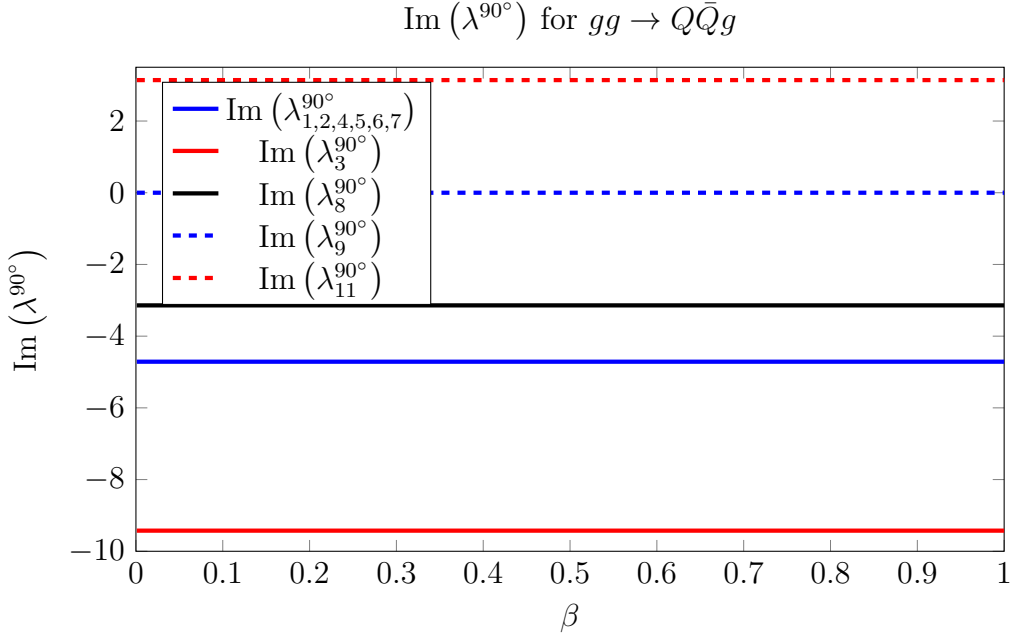
\begin{figure}
\centering
\begin{tikzpicture}
\begin{axis}[title=$\mathrm{Im}\left(\lambda^{90^{\circ}}\right)$ for $gg\rightarrow Q\bar{Q}g$,
xlabel={$\beta$},
ylabel={$\mathrm{Im}\left(\lambda^{90^{\circ}}\right)$},
ymin=-10, ymax=3.5,
xmin=0, xmax=1,
legend pos=north west,
legend entries={$\mathrm{Im}\left(\lambda^{90^{\circ}}_{1,2,4,5,6,7}\right)$,
$\mathrm{Im}\left(\lambda^{90^{\circ}}_{3}\right)$,
$\mathrm{Im}\left(\lambda^{90^{\circ}}_{8}\right)$,
$\mathrm{Im}\left(\lambda^{90^{\circ}}_{9}\right)$,
$\mathrm{Im}\left(\lambda^{90^{\circ}}_{11}\right)$},]
\addplot[blue, ultra thick] table {gluony90ima1.dat};
\addplot[red, ultra thick] table {gluony90im3.dat};
\addplot[black, ultra thick] table {gluony90im8.dat};
\addplot[blue, dashed, ultra thick] table {gluony90im9.dat};
\addplot[red, dashed, ultra thick] table {gluony90im11.dat};
\end{axis}
\end{tikzpicture}
\caption{The imaginary parts of the eigenvalues of $\tilde{\Gamma}_{S}$ for $gg\rightarrow Q\bar{Q}g$ at $\theta = 90^{\circ}$.}
\label{fig:gluonyim90}
\end{figure}

In this subsection, we perform an analogous analysis of the eigensystem for $gg\rightarrow Q\bar{Q}g$ to the case of $q\bar{q}\rightarrow Q\bar{Q}g$. The set of the eigenvalues is richer then in the scattering process of the quark and antiquark due to the larger colour basis. For $\theta = 90^{\circ}$ the real parts of the regularized eigenvalues are shown in fig. \ref{fig:gluonyre90} and the imaginary parts are shown in fig. \ref{fig:gluonyim90}. The singular matrix in $\beta_{34}$ has a form $\frac{i\pi}{\beta_{34}}\times\mathbf{diag}\left(-\frac{2}{3},-\frac{2}{3},\frac{1}{12},\frac{1}{12},\frac{1}{12},\frac{1}{12},\frac{1}{12},\frac{1}{12},\frac{1}{12},\frac{1}{12},\frac{1}{12}\right)$ in this case. One finds also one value of the leading small $\beta$ behaviour of the eigenvalues, which is the same as in the quark channel: \end{spacing}

\begin{equation}
\lambda^{\mathrm{sing}} = 6\log\beta.
\end{equation}

\begin{spacing}{2}

After the procedure of regularization (analogous to the $q\bar{q}$ scattering case) one can see some similarities for both the scattering angles. In the case $\theta = 90^{\circ}$ the eigensystem consists of 6 different eigenvalues. The degenerate eigenvalues are $\lambda_{1}=\lambda_{2}$, $\lambda_{3}=\lambda_{4}=\lambda_{5}=\lambda_{6}$ and $\lambda_{8}=\lambda_{9}$. The real parts of the eigenvalues are nearly constant up to $\beta\approx 0.6$ and all the imaginary parts are constant in whole range of $\beta$. The  results for $\theta = 30^{\circ}$ are shown in fig. \ref{fig:gluonyre30} (the real parts of eigenvalues) and in fig. \ref{fig:gluonyim30} (the imaginary parts of eigenvalues). The singular part of the eigenvalues at $\theta = 30^{\circ}$ is the same as for $\theta = 90^{\circ}$. The degeneracy of the eigensystem is lower (the degeneracy between eigenvalues 4,5,6,7 is reduced to the separate degeneracy $\lambda_{3}=\lambda_{4}$ and $\lambda_{5}=\lambda_{6}$). The real parts of eigenvalues are nearly flat for $\beta<0.5$ then they grow rapidly. $\mathrm{Im}(\lambda^{30^{\circ}}_{10})$ ($\mathrm{Im}(\lambda^{30^{\circ}}_{11})$) is a growing (decreasing) function of $\beta$. The imaginary parts of the remaining eigenvalues are constant.

Moreover, comparing the behaviour of the eigensystem of SAD matrices for  processes $q\bar{q}\rightarrow Q\bar{Q}g$ and $gg\rightarrow Q\bar{Q}g$ one  finds some similarities. For example, at $\theta = 90^{\circ}$ there is a constant behaviour in $\beta$ for the imaginary part of eigenvalues in both reactions. When the kinematic configuration becomes less symmetrical (the $\theta = 30^{\circ}$ case) the set of eigenvalues with a flat $\beta$-dependence is reduced.

\end{spacing}

\begin{figure}
\centering
\begin{tikzpicture}
\begin{axis}[title=$\mathrm{Re}\left(\lambda^{\mathrm{reg},30^{\circ}}\right)$ for $gg\rightarrow Q\bar{Q}g$,
xlabel={$\beta$},
ylabel={$\mathrm{Re}\left(\lambda^{\mathrm{reg},30^{\circ}}\right)$},
ymin=-6, ymax=4,
xmin=0, xmax=1,
legend pos=north west,
legend entries={$\mathrm{Re}\left(\lambda^{\mathrm{reg},30^{\circ}}_{1,2}\right)$,
$\mathrm{Re}\left(\lambda^{\mathrm{reg},30^{\circ}}_{3,8}\right)$,
$\mathrm{Re}\left(\lambda^{\mathrm{reg},30^{\circ}}_{4,9}\right)$,
$\mathrm{Re}\left(\lambda^{\mathrm{reg},30^{\circ}}_{5,10}\right)$,},]
\addplot[blue, ultra thick] table {gluonydrell30re1.dat};
\addplot[red, ultra thick] table {gluonydrell30re3.dat};
\addplot[black, dashed, ultra thick] table {gluonydrell30re4.dat};
\addplot[black, ultra thick] table {gluonydrell30re5.dat};
\end{axis}
\end{tikzpicture}
\\
\begin{tikzpicture}
\begin{axis}[title=$\mathrm{Re}\left(\lambda^{\mathrm{reg},30^{\circ}}\right)$ for $gg\rightarrow Q\bar{Q}g$,
xlabel={$\beta$},
ylabel={$\mathrm{Re}\left(\lambda^{\mathrm{reg},30^{\circ}}\right)$},
ymin=-11, ymax=8,
xmin=0, xmax=1,
legend pos=north west,
legend entries={$\mathrm{Re}\left(\lambda^{\mathrm{reg},30^{\circ}}_{6}\right)$,
$\mathrm{Re}\left(\lambda^{\mathrm{reg},30^{\circ}}_{7}\right)$,
$\mathrm{Re}\left(\lambda^{\mathrm{reg},30^{\circ}}_{11}\right)$},]
\addplot[blue, ultra thick] table {gluonydrell30re6.dat};
\addplot[red, ultra thick] table {gluonydrell30re7.dat};
\addplot[black, dashed, ultra thick] table {gluonydrell30re11.dat};
\end{axis}
\end{tikzpicture}
\caption{The real parts of the regularized eigenvalues of $\tilde{\Gamma}_{S}$ for $gg\rightarrow Q\bar{Q}g$ \qquad \qquad at $\theta = 30^{\circ}$.}
\label{fig:gluonyre30}
\end{figure}
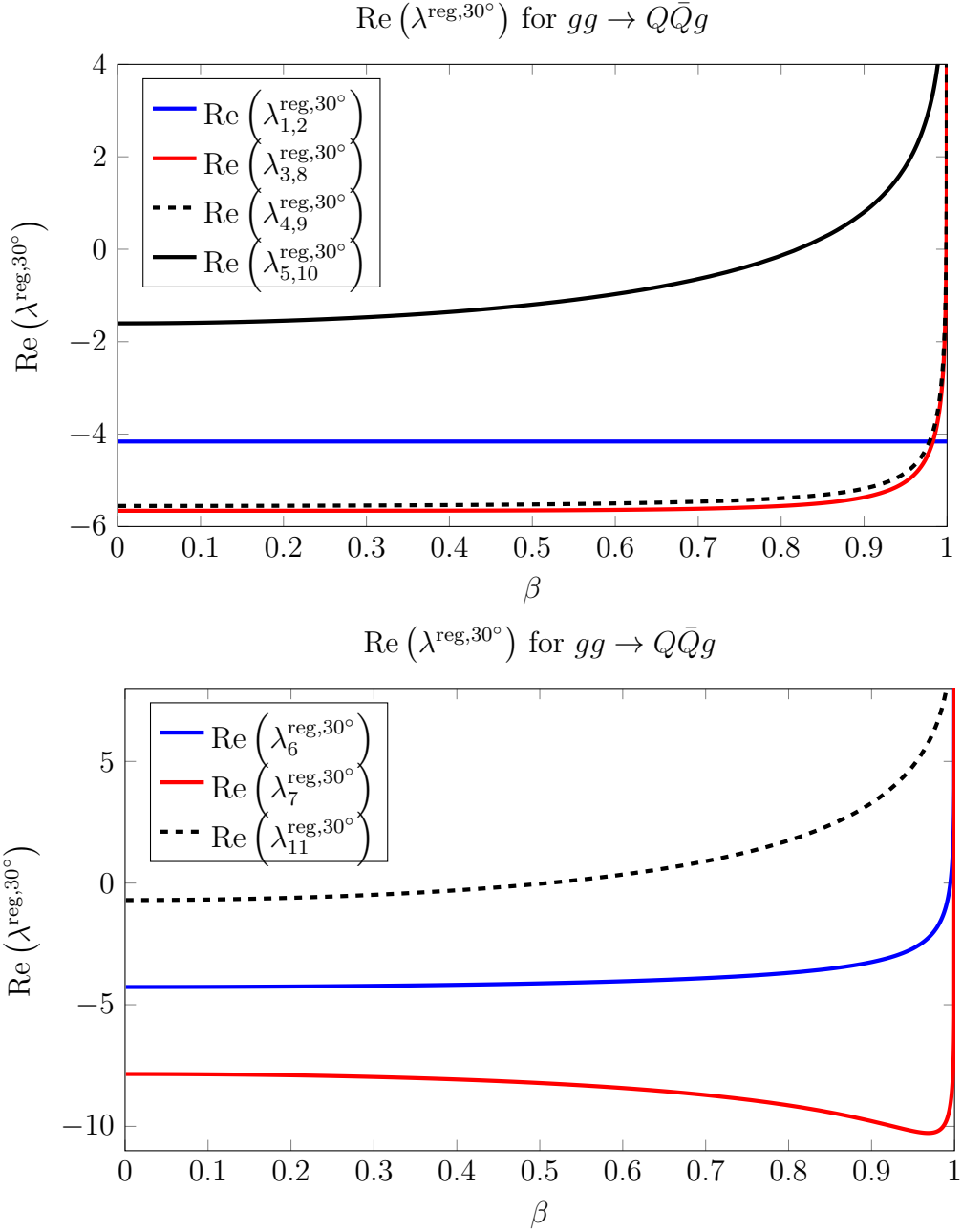

\begin{figure}
\centering
\begin{tikzpicture}
\begin{axis}[title=$\mathrm{Im}\left(\lambda^{30^{\circ}}\right)$ for $gg\rightarrow Q\bar{Q}g$,
xlabel={$\beta$},
ylabel={$\mathrm{Im}\left(\lambda^{30^{\circ}}\right)$},
ymin=-8.1, ymax=2,
xmin=0, xmax=1,
legend pos=north west,
legend entries={$\mathrm{Im}\left(\lambda^{30^{\circ}}_{1,2}\right)$,
$\mathrm{Im}\left(\lambda^{30^{\circ}}_{3,4,5,6}\right)$,
$\mathrm{Im}\left(\lambda^{30^{\circ}}_{7}\right)$,
$\mathrm{Im}\left(\lambda^{30^{\circ}}_{8,9}\right)$,
$\mathrm{Im}\left(\lambda^{30^{\circ}}_{10}\right)$,
$\mathrm{Im}\left(\lambda^{30^{\circ}}_{11}\right)$},]
\addplot[blue, ultra thick] table {gluony30ima1.dat};
\addplot[red, ultra thick] table {gluony30im3.dat};
\addplot[black, dashed, ultra thick] table {gluony30im7.dat};
\addplot[black, ultra thick] table {gluony30im8.dat};
\addplot[blue, dashed, ultra thick] table {gluony30im10.dat};
\addplot[red, dashed, ultra thick] table {gluony30im11.dat};
\end{axis}
\end{tikzpicture}
\caption{The imaginary parts of the eigenvalues of $\tilde{\Gamma}_{S}$ for $gg\rightarrow Q\bar{Q}g$ at $\theta = 30^{\circ}$.}
\label{fig:gluonyim30}
\end{figure}
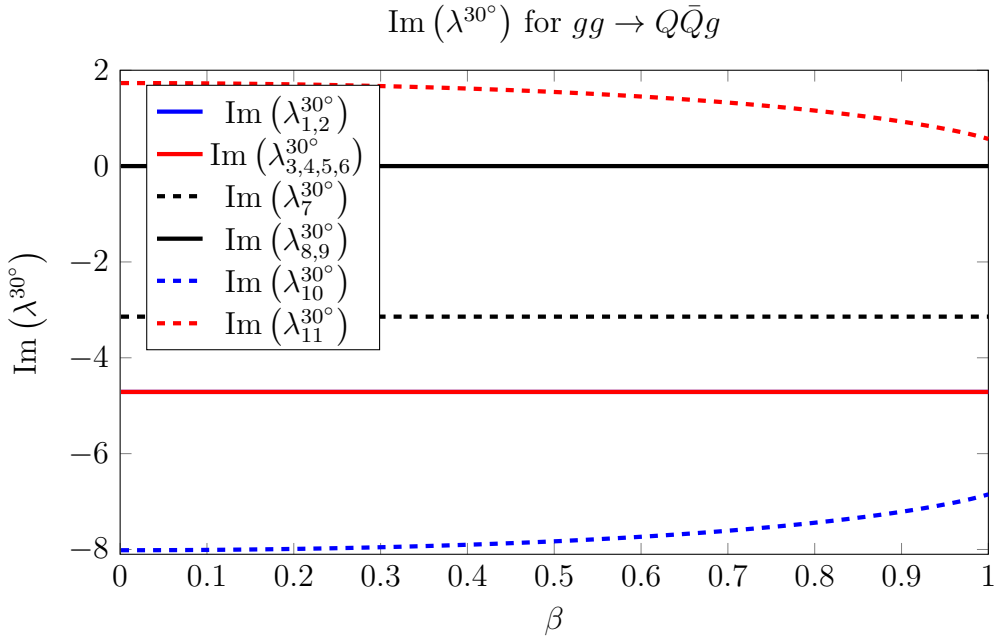

\begin{spacing}{2}

\section{Discussion and summary}

In this paragraph we compare the calculated regularized eigenvalues $\lambda^{\mathrm{reg}}_{i}$ of the SAD matrices to the SAD eigenvalues for processes $q\bar{q}\rightarrow Q\bar{Q}$ and $gg\rightarrow Q\bar{Q}$ in the small $\beta$ region. Note that the full eigenvalues for $2\to 3$ processes contain in addition to the regular parts a negative singular term $6\log\beta$ for the $q\bar{q}$ and $gg$ channel. The logarithmic terms combine with the dominant regular terms into even larger negative terms of the eigenvalues, that is they lead to stronger effects of gluon radiation. For the process $q\bar{q}\rightarrow Q\bar{Q}$ the real parts of the two at $\beta\rightarrow 0$ tend to $-1.5\frac{\alpha_{s}}{\pi}$ and $0$. Recall that for the case of $q\bar{q}\rightarrow Q\bar{Q}g$ the largest (negative) SAD eigenvalue reads $\mathrm{Re}\left(\lambda^{\mathrm{reg},90^{\circ}}\right)=-2.5\frac{\alpha_{s}}{\pi}$ (for $\theta=90^{\circ}$) and $\mathrm{Re}\left(\lambda^{\mathrm{reg},30^{\circ}}\right)=-5\frac{\alpha_{s}}{\pi}$ (for $\theta=30^{\circ}$). It means that the effect of soft gluon radiation for $q\bar{q}\rightarrow Q\bar{Q}g$ is almost two times stronger (the $90^{\circ}$ case) or three times larger (the $30^{\circ}$ case). In the gluonic case the radiation effects are even stronger. For $gg\rightarrow Q\bar{Q}g$ we obtained $\mathrm{Re}\left(\lambda^{\mathrm{reg},90^{\circ}}\right)=-4\frac{\alpha_{s}}{\pi}$ (for $\theta=90^{\circ}$) and $\mathrm{Re}\left(\lambda^{\mathrm{reg},30^{\circ}}\right)=-8\frac{\alpha_{s}}{\pi}$ (for $\theta=30^{\circ}$), so the radiation is enhanced by factors three and five correspondingly with respect to the process $gg\rightarrow Q\bar{Q}$. The imaginary parts of eigenvalues cancel out in the regime $\beta\rightarrow 0$, so we will not discuss them. These results imply that the soft gluon radiation is a source of enhanced corrections for the heavy quark pair production in association with a gluon jet.

In this paper we have derived the one-loop soft anomalous dimension matrices for $q\bar{q}\rightarrow Q\bar{Q}g$ and $gg\rightarrow Q\bar{Q}g$. We presented the SAD matrices for an arbitrary scattering angle $\theta$ of a clustered pair of heavy quark and antiquark with respect to the incoming parton axis in the CMS frame. We also analyzed the spectrum of the eigenvalues of the SAD matrices in details for two kinematic configurations $\theta = 90^{\circ}$ and $30^{\circ}$, performing explicit numerical calculations of the SAD eigenvalues. The obtained results are a step towards implementing the soft resummation procedure for $Q\bar{Q}$ - jet production in hadron colliders, and improving accuracy of theoretical predictions.

\section*{Acknowledgments}

The author would like to thank Prof. L. Motyka for the help of preparing this paper, Prof. M. Praszałowicz for valuable comments on the manuscript, Prof. A. Kulesza for the discussion, Prof. M. Sj\"odahl and R. Sch\"afer for the correspondence. This work was supported by the Polish NCN grant DEC-2014/13/B/ST2/02486. 

\end{spacing}

\end{document}